# Algorithms for integrals of holonomic functions over domains defined by polynomial inequalities

Toshinori Oaku

Department of Mathematics, Tokyo Woman's Christian University

October 28, 2011

**Abstract**

A holonomic function is a differentiable or generalized function which satisfies a holonomic system of linear partial or ordinary differential equations with polynomial coefficients. The main purpose of this paper is to present algorithms for computing a holonomic system for the definite integral of a holonomic function with parameters over a domain defined by polynomial inequalities. If the integrand satisfies a holonomic difference-differential system including parameters, then a holonomic difference-differential system for the integral can also be computed. In the algorithms, holonomic distributions (generalized functions in the sense of L. Schwartz) are inevitably involved even if the integrand is a usual function.

## Introduction

Holonomic systems of linear differential equations, which play a central role in the $D$-module theory, were introduced by Bernstein [2] in the algebraic setting, and by Sato et al. [20] in the analytic setting under the name of 'maximally overdetermined systems'. We follow the formulation by Bernstein, which would be the more adapted to practical applications with computers. Hence, in the present paper, we mean by a *holonomic function* a function which satisfies a holonomic system of linear differential equations with polynomial coefficients. Two equivalent definitions of a holonomic system will be recalled in Section 1.

Most of the special functions in one variable such as various hypergeometric functions and the Bessel function are holonomic by the definition. As an important class of holonomic functions in several variables, let us consider the multi-valued analytic function $u = f_1^{\lambda_1} \cdots f_m^{\lambda_m}$ with non-zero polynomials $f_1, \ldots, f_m$ in $n$ variables. As a multi-valued analytic function, $u$ is defined on $\{x \in \mathbb{C}^n \mid f_1(x) \cdots f_m(x) \neq 0\}$ and is holonomic for any complex number $\lambda_j$. We can regard this function also as a distribution $(f_1)_+^{\lambda_1} \cdots (f_m)_+^{\lambda_m}$ defined on $\mathbb{R}^n$ in the sense of L. Schwartz if $f_j$ are real polynomials and $(\lambda_1, \ldots, \lambda_m)$ avoids some exceptional set. Such a distribution was introduced by Gel'fand and Shilov [7] in some restricted cases. See e.g., [10] for a theoretical study on generalized functions including such a distribution. In particular, substituting zeros for the parameters yields

$$(f_1)_+^0 \cdots (f_m)_+^0 = Y(f_1) \cdots Y(f_m),$$

where $Y$ denotes the Heaviside function, i.e., $Y(t) = 1$ for $t > 0$ and $Y(t) = 0$ for $t \leq 0$. The Heaviside function will play an essential role in the present paper.

Algorithmic studies, especially the integration, of holonomic functions were pioneered by Almkvist and Zeilberger [1] who introduced what is called the creative telescoping method which applies to both difference and differential holonomic systems. Then by using Gröbner bases in the ring of differential operators, Takayama ([23], [24]) introduced two algorithms for integration of holonomic functions. These algorithms were generalized to Ore algebras by Chyzak [5] (see also [6]).

On the other hand, a precisely $D$-module theoretic algorithm was given in [18] (see also [21], [17]), which is the 'Fourier transform' of the restriction algorithm firstly introduced by Oaku [15] and generalized in [19]. Given a holonomic system for a function $u(x_1, \ldots, x_n)$, this algorithm outputs a holonomic system for the integral

$$v(x_1, \ldots, x_{n-d}) = \int_{\mathbb{R}^d} u(x_1, \ldots, x_n) \, dx_{n-d+1} \cdots dx_n.$$

This algorithm might not be efficient enough but has an advantage in the following two respects: First, the holonomicity of the output is guaranteed in the $D$-module theory if the input is holonomic. Second, it applies to generalized functions as well as to infinitely differentiable functions since it does not involve computation with rational functions as coefficients.

The purpose of the present paper is to apply this $D$-module theoretic algorithm to the integral over a domain defined by one or more polynomial inequalities by using the Heaviside function, generalizing and elaborating a method sketched in [17]. More precisely, we are concerned with an integral

$$v(x_1, \ldots, x_{n-d}) = \int_{D(x_1, \ldots, x_{n-d})} u(x_1, \ldots, x_n) \, dx_{n-d+1} \cdots dx_n \qquad (1)$$

over a domain

$$D(x_1, \ldots, x_{n-d}) = \{(x_{n-d+1}, \ldots, x_n) \in \mathbb{R}^d \mid f_1(x_1, \ldots, x_n) \geq 0, \ldots, f_m(x_1, \ldots, x_n) \geq 0\}$$

with polynomials $f_1, \ldots, f_m$ with real coefficients.

Let us present two explanatory examples.

**Example 1** Set
$$v(x) = \int_0^1 e^{xy} \, dy = \frac{e^x - 1}{x}.$$

The integrand $u(x, y) = e^{xy}$ satisfies a holonomic system
$$(\partial_x - y)u(x, y) = (\partial_y - x)u(x, y) = 0$$

with $\partial_x = \partial/\partial x$ and $\partial_y = \partial/\partial y$. If we apply the integration algorithm to this holonomic system, we get an incorrect equation $xv(x) = 0$, which follows from formal integration of the equation $(\partial_y - x)e^{xy} = 0$ with respect to $y$; this integral does not exist in fact.

In order to get rid of the boundary conditions at $y = 0, 1$, we rewrite this integral in terms of the Heaviside function $Y(t)$ to get

$$v(x) = \int_{-\infty}^{\infty} e^{xy} Y(y) Y(1-y) \, dy.$$

The new integrand $\tilde{u}(x,y) = e^{xy}Y(y)Y(1-y)$ satisfies a holonomic system

$$y(y-1)(\partial_y - x)\tilde{u}(x,y) = (\partial_x - y)\tilde{u}(x,y) = 0$$

in the sense of distribution theory. In fact one has

$$y(y-1)(\partial_y - x)(e^{xy}Y(y)Y(1-y)) = y(y-1)e^{xy}(\delta(y) - \delta(y-1)) = 0,$$

where $\delta(y) = \partial_y Y(y)$ denotes the Dirac delta function. The integration algorithm applied to this holonomic system outputs an answer

$$(x\partial_x^2 - (x-2)\partial_x - 1)v(x) = 0,$$

which is correct as is seen by rewriting the operator as follows:

$$x\partial_x^2 - (x-2)\partial_x - 1 = \partial_x(\partial_x - 1)x.$$

Of course, one can treat integrals like this one over an interval by the classical creative telescoping method taking the boundary conditions into account (see e.g, [5]). See also a recent work by [11] for an algorithm to compute inhomogeneous differential equations for such an integral. However, it does not seem straightforward to apply these methods to more general integrals such as the following:

**Example 2** Set

$$v(t) = \int_{x^2+y^2 \leq t} \frac{dxdy}{1+x^2+y^2} = \int_{\mathbb{R}^2} (1+x^2+y^2)^{-1}Y(t-x^2-y^2)\,dxdy.$$

Note that $v(t)$ is continuous in $t \in \mathbb{R}$ and $v(t) = 0$ holds for $t < 0$. By using an algorithm in [18], we get generators

$$\begin{aligned}
& y\partial_x - x\partial_y, \quad (-x^2 - y^2 + t)\partial_t - s_1, \\
& ((-x^2 + t)\partial_t - s_1)\partial_x - yx\partial_t\partial_y - 2x\partial_t, \\
& (-x^2 - y^2 - 1)\partial_y + (-2t-2)y\partial_t + (2s_1 + 2s_2)y, \\
& (x^2 + 1)\partial_x + yx\partial_y + (2t+2)x\partial_t + (-2s_1 - 2s_2)x, \\
& ((-t-1)x\partial_t + s_1 x)\partial_x + ((-t-1)y\partial_t + s_1 y)\partial_y + (-2t^2 - 2t)\partial_t^2 \\
& \quad + ((4s_1 + 2s_2 - 2)t + 2s_1 - 2)\partial_t - 2s_1^2 - 2s_2 s_1
\end{aligned}$$

of the annihilating ideal of the analytic function $(t-x^2-y^2)^{s_1}(1+x^2+y^2)^{s_2}$ with parameters $s_1, s_2$. Substituting 0 for $s_1$ and $-1$ for $s_2$ gives a holonomic system for the distribution

$$u(x,y,t) = (1+x^2+y^2)^{-1}Y(t-x^2-y^2).$$

Then applying the integration algorithm to this holonomic system, we obtain a differential equation

$$((t^2+t)\partial_t^2 + t\partial_t)v = 0.$$

In fact, integration in polar coordinates gives $v(t) = \pi Y(t)\log(1+t)$. It is easy to verify that $v(t)$ satisfies the above equation as distribution.

For the general integral (1), we rewrite it in the form

$$v(x_1,\ldots,x_{n-d}) = \int_{\mathbb{R}^d} u(x_1,\ldots,x_n) Y(f_1(x_1,\ldots,x_n)) \cdots Y(f_m(x_1,\ldots,x_n))\, dx_{n-d+1} \cdots dx_n.$$

Hence the integration algorithm can be applied if a holonomic system for the product $uY(f_1)\ldots Y(f_m)$ is computed. We shall describe the procedure in detail and prove its validity. We remark that the 'indefinite integral'

$$v(x_1,\ldots,x_n) = \int_a^{x_n} u(x_1,\ldots,x_{n-1},t)\, dt$$

is a special case of (1).

Our algorithm consists of the following two steps:

1. For a given holonomic system for a function $u$, compute a holonomic system for $uY(f_1)\cdots Y(f_m)$.

2. Compute a holonomic system for the integral $\int_{\mathbb{R}^d} uY(f_1)\cdots Y(f_m)\, dx_{n-d+1}\cdots dx_n$.

For the first step, we begin with an algorithm to compute a holonomic system for $Y(f_1)\cdots Y(f_m)$. Then the tensor product computation for $D$-modules gives an answer to the first step. For the case where $u$ is a complex power, or an exponential of a polynomial, the tensor product computation is unnecessary as in the previous examples. Even for general $u$, we shall give an alternative method which avoids the tensor product computation. The second step can be done by the integration algorithm for $D$-modules.

In many practical examples, the integrand can have auxiliary parameters other than $x_1,\ldots,x_n$ above but cannot be regarded as a holonomic function including the parameters. For example, consider the integral

$$v(x_1,\ldots,x_{n-d},s_1,\ldots,s_m) = \int_{D(x_1,\ldots,x_{n-d})} u(f_1)_+^{s_1}\cdots(f_m)_+^{s_m}\, dx_{n-d+1}\cdots dx_n$$

with parameters $s_1,\ldots,s_m$. The integration algorithm cannot be applied directly unless we specify the values of the parameters $s_1,\ldots,s_m$ explicitly. However, it is often the case, as with the example above, that the integrand satisfies a holonomic system of difference-differential equations including the parameters. We also give an algorithm for computing a holonomic difference-differential system for such an integral.

We have implemented the algorithms in the present paper by using a computer algebra system Risa/Asir [12]. In particular, we make use of a library file `nk_restriction` by H. Nakayama (see [11]) for computing restriction and integration.

## 1 Differential operators and holonomic systems

Let us denote by $D_n$ the ring of differential operators on the variables $x = (x_1,\ldots,x_n)$ with polynomial coefficients. An element $P$ of $D_n$ is written in a finite sum

$$P = \sum_{\alpha,\beta\in\mathbb{N}^n} a_{\alpha,\beta} x^\alpha \partial^\beta, \tag{2}$$

where $\alpha = (\alpha_1, \ldots, \alpha_n)$, $\beta = (\beta_1, \ldots, \beta_n) \in \mathbb{N}^n$ are vectors of nonnegative integers with $\mathbb{N} = \{0, 1, 2, \ldots\}$, $x^\alpha = x_1^{\alpha_1} \cdots x_n^{\alpha_n}$, $\partial^\beta = \partial_1^{\beta_1} \cdots \partial_n^{\beta_n}$ with the derivations $\partial_i = \partial/\partial x_i$ ($i = 1, \ldots, n$), and $a_{\alpha, \beta}$ are complex numbers.

Given $P_1, \ldots, P_r \in D_n$, we associate the left ideal $I = D_n P_1 + \cdots + D_n P_r$ generated by $P_1, \ldots, P_r$ with a system of linear differential equations

$$P_1 u = \cdots = P_r u = 0 \tag{3}$$

for an unknown function $u$. This enables us to work with a left ideal of $D_n$ instead of each system of linear differential equations. Here we suppose that the unknown function $u$ belongs to a 'function space' $\mathcal{F}$ which is a left $D_n$-module. Examples of such $\mathcal{F}$ are the set $C^\infty(U)$ of $C^\infty$ functions on an open subset $U$ of $\mathbb{R}^n$, the set $\tilde{\mathcal{O}}(U)$ of possibly multi-valued analytic functions on an open subset $U$ of $\mathbb{C}^n$, the set $\mathcal{D}'(U)$ of the Schwartz distributions on an open subset $U$ of $\mathbb{R}^n$, and the set $S'(\mathbb{R}^n)$ of tempered distributions.

A weight vector for $D_n$ is an integer vector

$$w = (w_1, \ldots, w_n; w_{n+1}, \cdots w_{2n}) \in \mathbb{Z}^{2n}$$

with the conditions $w_i + w_{n+i} \geq 0$ for $i = 1, \ldots, n$, which are necessary in view of the commutation relation $\partial_i x_i = x_i \partial_i + 1$ in $D_n$. For a nonzero differential operator $P$ of the form (2), we define its $w$-*order* to be

$$\mathrm{ord}_w(P) := \max\{\langle w, (\alpha, \beta)\rangle = w_1 \alpha_1 + \cdots + w_n \alpha_n + w_{n+1} \beta_1 + \cdots + w_{2n} \beta_n \mid a_{\alpha, \beta} \neq 0\}.$$

We set $\mathrm{ord}_w(0) := -\infty$. A weight vector $w$ induces the $w$-*filtration*

$$F_k^w(D_n) := \{P \in D_n \mid \mathrm{ord}_w(P) \leq k\} \quad (k \in \mathbb{Z})$$

on the ring $D_n$. Following Bernstein [2], let us define the notion of holonomic system by using the weight vector $(\mathbf{1}, \mathbf{1}) = (1, \ldots, ; 1, \ldots, 1) \in \mathbb{Z}^{2n}$. Let $M$ be a left $D_n$-module and $\{F_k(M)\}_{k \in \mathbb{Z}}$ be a good $(\mathbf{1}, \mathbf{1})$-filtration. This means the following properties:

1. every $F_k(M)$ is a finite dimensional vector space over $\mathbb{C}$;

2. $F_k(M) \subset F_{k+1}(M)$ for all $k \in \mathbb{Z}$;

3. $\bigcup_{k \in \mathbb{Z}} F_k(M) = M$;

4. $F_i^{(\mathbf{1},\mathbf{1})}(D_n) F_k(M) \subset F_{i+k}(M)$ for all $i, k \in \mathbb{Z}$;

5. there exists $k_1 \in \mathbb{Z}$ such that $F_k(M) = 0$ for $k \leq k_1$;

6. there exists $k_2 \in \mathbb{Z}$ such that $F_i^{(\mathbf{1},\mathbf{1})}(D_n) F_k(M) = F_{i+k}(M)$ for $k \geq k_2$.

Then there exists a polynomial in $k$ such that $\dim_\mathbb{C} F_k(M) = p(k)$ for sufficiently large $k$. The degree of $p(k)$ does not depend on the choice of a good $(\mathbf{1}, \mathbf{1})$-filtration of $M$ and is called the dimension of the module $M$, which we denote by $d(M)$. It was proved by Bernstein [2] that $d(M) \geq n$ if $M \neq 0$. Following Bernstein let us adopt the following

**Definition 1** A finitely generated left $D_n$-module $M$ is called a *holonomic system* if $d(M) \leq n$. We also call a left ideal $I$ of $D_n$ to be a *holonomic ideal*, by abuse of terminology, if the left $D_n$-module $D_n/I$ is holonomic.

Note that $d(M) \leq n$ is equivalent to $d(M) = n$ or $M = 0$ in view of the Bernstein inequality stated above. The dimension $d(M)$ can be computed as the degree of the Hilbert function from a Gröbner base with respect to a term order which is compatible with the total degree.

Let us recall another characterization of a holonomic system essentially given by Sato et al. [20]. For this we use the weight vector $w = (\mathbf{0}, \mathbf{1}) = (0, \ldots, 0, 1, \ldots, 1)$. Let $P$ be a nonzero differential operator written in the form (2) and set $m := \mathrm{ord}_{(\mathbf{0},\mathbf{1})}(P)$. Then the *principal symbol* of $P$ is the polynomial defined by

$$\sigma(P)(x, \xi) = \sum_{|\beta|=m} \sum_{\alpha} a_{\alpha,\beta} x^\alpha \xi^\beta,$$

where $\xi = (\xi_1, \ldots, \xi_n)$ are the commutative variables corresponding to the derivations $\partial = (\partial_1, \ldots, \partial_n)$.

Set $M := D_n/I$ with a left ideal $I$ of $D_n$. The *characteristic variety* of $M$ is defined to be the algebraic set

$$\mathrm{Char}(M) := \{(x, \xi) \in \mathbb{C}^{2n} \mid \sigma(P)(x, \xi) = 0 \text{ for any } P \in I \setminus \{0\}\}$$

of $\mathbb{C}^{2n}$. It was proved in [20] that the dimension of (every irreducible component of) $\mathrm{Char}(M)$ is not less than $n$ unless $M = 0$. Especially, $M$ is holonomic if and only if the dimension of the characteristic variety is $n$ or else $M = 0$. The equivalence of these two definitions of holonomic system is proved, e.g. in Chapter 3 of [3]. The characteristic variety can be computed via a Gröbner base with respect to a term order which is compatible with the $(\mathbf{0}, \mathbf{1})$-weight (cf. [13]).

## 2 Holonomic distributions

First let us recall the definition of distributions due to [22].

**Definition 2** Let $C_0^\infty(U)$ be the set of the $C^\infty$ functions on an open set $U$ of $\mathbb{R}^n$ with compact support. A distribution $u$ on $U$ is a linear functional

$$u : C_0^\infty(U) \ni \varphi \longmapsto \langle u, \varphi \rangle \in \mathbb{C}$$

such that $\lim_{j \to \infty} \langle u, \varphi_j \rangle = 0$ holds for a sequence $\{\varphi_j\}$ of $C_0^\infty(U)$ if there is a compact set $K$ contained in $U$ such that $\varphi_j$ are zero on $U \setminus K$ and

$$\lim_{j \to \infty} \sup_{x \in U} |\partial^\alpha \varphi_j(x)| = 0 \quad \text{for any } \alpha \in \mathbb{N}^n.$$

The set of the distributions on $U$ is denoted by $\mathcal{D}'(U)$. The derivative $\partial_k u$ of $u$ with respect to $x_k$ is defined by

$$\langle \partial_k u, \varphi \rangle = -\langle u, \partial_k \varphi \rangle \quad \text{for any } \varphi \in C_0^\infty(U).$$

For a $C^\infty$ function $a$ on $U$, the product $au$ is defined by

$$\langle au, \varphi \rangle = \langle u, a\varphi \rangle \quad \text{for any } \varphi \in C_0^\infty(U).$$

In particular, by these actions of the derivations and the polynomial multiplications, $\mathcal{D}'(U)$ has a natural structure of left $D_n$-module.

**Definition 3** Let $u$ be a $C^\infty$ function or a distribution defined on an open subset $U$ of $\mathbb{R}^n$. Then we call $u$ a *holonomic function* or a *holonomic distribution* on $U$ if $u$ satisfies a holonomic system. In other words, $u$ is holonomic if and only if its *annihilator*

$$\mathrm{Ann}_{D_n} u := \{P \in D_n \mid Pu = 0 \text{ on } U\}$$

is a holonomic ideal.

Let $f_1, \ldots, f_m$ be polynomials with real coefficients. We assume that the set $\{x \in \mathbb{R}^n \mid f_i(x) > 0 \, (i = 1, \ldots, m)\}$ is not empty. Then the distribution $u = (f_1)_+^{\lambda_1} \cdots (f_m)_+^{\lambda_m}$ on $\mathbb{R}^n$ is defined to be

$$\langle u, \varphi \rangle = \int_{f_1 \geq 0, \ldots, f_m \geq 0} f_1(x)^{\lambda_1} \cdots f_m(x)^{\lambda_m} \varphi(x) \, dx$$

for $\varphi \in C_0^\infty(\mathbb{R}^n)$ if $\mathrm{Re}\, \lambda_i \geq 0$ for each $i$. Moreover, $u$, that is, $\langle u, \varphi \rangle$ for any $\varphi \in C_0^\infty(\mathbb{R}^n)$, is holomorphic in $(\lambda_1, \ldots, \lambda_m)$ on the domain

$$\Omega_+ := \{(\lambda_1, \ldots, \lambda_m) \in \mathbb{C}^m \mid \mathrm{Re}\, \lambda_i > 0 \quad (i = 1, \ldots, m)\}$$

and is continuous in $(\lambda_1, \ldots, \lambda_m)$ on the closure of $\Omega_+$.

In order to deduce a holonomic system for the distribution $(f_1)_+^{\lambda_1} \cdots (f_m)_+^{\lambda_m}$, let us first work with the 'formal' function $f_1^{s_1} \cdots f_m^{s_m}$. More precisely, setting $F = f_1 \cdots f_m$, we consider a free module

$$\mathcal{L} := \mathbb{C}[x, s, F^{-1}] f_1^{s_1} \cdots f_m^{s_m}$$

over $\mathbb{C}[x, s, F^{-1}]$, which has also a natural structure of left $D_n[s]$-module induced from the formal derivation

$$\partial_i (a f_1^{s_1} \cdots f_m^{s_m}) = \frac{\partial a}{\partial x_i} f_1^{s_1} \cdots f_m^{s_m} + a \sum_{j=1}^m s_j \frac{\partial f_j}{\partial x_i} f_j^{-1} f_1^{s_1} \cdots f_m^{s_m}$$

with $a \in \mathbb{C}[x, s, F^{-1}]$. Then the annihilator ideal

$$\mathrm{Ann}_{D_n[s]} f_1^{s_1} \cdots f_m^{s_m} = \{P(s) \in D_n[s] \mid P(s) f_1^{s_1} \cdots f_m^{s_m} = 0 \text{ in } \mathcal{L}\}$$

can be computed precisely with an algorithm of [18], or of [4] if the coefficients of $f_1, \ldots, f_m$ are contained in a computable subfield of $\mathbb{C}$.

Define a $D_n[s]$-submodule $\mathcal{N}$ of $\mathcal{L}$ by

$$\mathcal{N} := D_n[s] f_1^{s_1} \cdots f_m^{s_m} \simeq D_n[s]/\mathrm{Ann}_{D_n[s]} f_1^{s_1} \cdots f_m^{s_m}.$$

Let us specialize the parameters $s$. For $\lambda = (\lambda_1, \ldots, \lambda_m) \in \mathbb{C}^m$, set

$$\mathcal{N}(\lambda) := \mathcal{N}/((s_1 - \lambda_1)\mathcal{N} + \cdots + (s_m - \lambda_m)\mathcal{N}).$$

Then $\mathcal{N}(\lambda)$ has a natural structure of left $D_n$-module which is induced from the left $D_n[s]$-module structure of $\mathcal{N}$ by the identification

$$D_n \simeq D_n[s]/((s_1 - \lambda_1)D_n[s] + \cdots + (s_m - \lambda_m)D_n[s]).$$

**Proposition 1** $\mathcal{N}(\lambda)$ *is a holonomic $D_n$-module for any $\lambda \in \mathbb{C}^n$.*

Proof: We use a filtration

$$F_i(D_n[s]) := \{P(s) = \sum_{j=0}^{k} P_j s^j \mid P_j \in D_n, \, j + \mathrm{ord}_{(\mathbf{1},\mathbf{1})}(P_j) \leq i, \, k \in \mathbb{N}\} \quad (i \in \mathbb{Z})$$

on $D_n[s]$. Let us define a filtration

$$F_k(\mathcal{L}) := \left\{ \frac{a}{F^k} f_1^{s_1} \cdots f_m^{s_m} \mid a \in \mathbb{C}[x,s], \, \deg a \leq dk \right\} \quad (k \in \mathbb{N})$$

on $\mathcal{L}$, where $\deg$ denotes the total degree in $(x,s)$ and $d = \deg F + 1$. Then it is easy to see that

$$F_k(\mathcal{N}) := F_k(D_n[s]) f_1^{s_1} \cdots f_m^{s_m} \subset F_k(\mathcal{L})$$

holds for $k \geq 0$, which implies an inequality

$$\dim_{\mathbb{C}} F_k(\mathcal{N}) \leq \dim_{\mathbb{C}} F_k(\mathcal{L}) = \binom{dk + n + m}{n + m}.$$

There exists a polynomial $p(k)$ in $k$ such that $\dim_{\mathbb{C}} F_k(\mathcal{N}) = p(k)$ holds for integers $k$ large enough since $\{F_k(\mathcal{N})\}$ is a good filtration on $\mathcal{N}$. From the above inequality, it follows that the degree of $p(k)$ is at most $n + m$.

Set $\mathcal{N}'(\lambda_m) := \mathcal{N}/(s_m - \lambda_m)\mathcal{N}$ and define a good filtration on it by

$$F_k(\mathcal{N}'(\lambda_m)) = F_k(D_n[s]) f^s / (F_k(D_n[s]) f^s \cap (s_m - \lambda_m)\mathcal{N}) \quad (k \in \mathbb{N})$$

with $f^s := f_1^{s_1} \cdots f_m^{s_m}$. Since $(s_m - \lambda_m) F_{k-1}(D_n[s]) f^s$ is contained in $F_k(D_n[s]) f^s \cap (s_m - \lambda_m)\mathcal{N}$ and the homomorphism $s_m - \lambda_m : \mathcal{N} \to \mathcal{N}$ is injective in view of the definition of $\mathcal{N}$, we have

$$\begin{aligned} \dim_{\mathbb{C}} F_k(\mathcal{N}'(\lambda_m)) &= \dim_{\mathbb{C}} F_k(D_n[s]) f^s - \dim_{\mathbb{C}} (F_k(D_n[s]) f^s \cap (s_m - \lambda_m)\mathcal{N}) \\ &\leq \dim_{\mathbb{C}} F_k(D_n[s]) f^s - \dim_{\mathbb{C}} F_{k-1}(D_n[s]) f^s \\ &= p(k) - p(k-1) \end{aligned}$$

for $k$ sufficiently large. It follows that $\dim_{\mathbb{C}} F_k(\mathcal{N}'(\lambda_m))$ coincides with a polynomial in $k$ of degree $\leq n + m - 1$ for sufficiently large $k$. Proceeding in the same way, we can show that $\dim_{\mathbb{C}} F_k(\mathcal{N}(\lambda))$ is a polynomial in $k$ of degree $\leq n$ for $k$ sufficiently large with

$$F_k(\mathcal{N}(\lambda)) := F_k(\mathcal{N})/(F_k(\mathcal{N}) \cap ((s_1 - \lambda_1)\mathcal{N} + \cdots + (s_m - \lambda_m)\mathcal{N})).$$

Note that we have

$$F_k(\mathcal{N}(\lambda)) = F_k^{(\mathbf{1},\mathbf{1})}(D_n)[f_1^{s_1} \cdots f_m^{s_m}]$$

with $[f_1^{s_1} \cdots f_m^{s_m}]$ being the modulo class of $f_1^{s_1} \cdots f_m^{s_m}$ in $\mathcal{N}(\lambda)$. This implies that $\mathcal{N}(\lambda)$ is a holonomic $D_n$-module. $\square$

Now let us make explicit the relation between the algebraic module $\mathcal{N}(\lambda)$ and the distribution $(f_1)_+^{\lambda_1} \cdots (f_m)_+^{\lambda_m}$.

**Lemma 1** Let $P(s) = P(s_1, \ldots, s_m)$ be an element of $D_n[s]$. Then
$$P(\lambda_1, \ldots, \lambda_m)(f_1)_+^{\lambda_1} \cdots (f_m)_+^{\lambda_m} = 0$$
holds in $\mathcal{D}'(\mathbb{R}^n)$ for any $(\lambda_1, \ldots, \lambda_m) \in \Omega_+$ if and only if $P(s)f_1^{s_1} \cdots f_m^{s_m} = 0$ holds in $\mathcal{L}$.

Proof: It is easy to see that
$$\partial_i((f_1)_+^{\lambda_1} \cdots (f_m)_+^{\lambda_m}) = \sum_{j=1}^{m} \lambda_j \frac{\partial f_j}{\partial x_i}(f_1)_+^{\lambda_1} \cdots (f_j)_+^{\lambda_j - 1} \cdots (f_m)_+^{\lambda_m}$$
holds as distribution and the right-hand side is a locally integrable function if Re $\lambda_j > 1$ for $j = 1, \ldots, m$, in accordance with the action of $\partial_i$ on $\mathcal{L}$. Let $P(s) \in D_n[s]$. There are $a(x, s) \in \mathbb{C}[x, s]$ and a positive integer $k$ such that
$$P(s)f_1^{s_1} \cdots f_m^{s_m} = \frac{a(x, s)}{(f_1 \cdots f_m)^k} f_1^{s_1} \cdots f_m^{s_m}$$
holds in $\mathcal{L}$. Moreover, the right-hand side of this equality vanishes in $\mathcal{L}$ if and only if $a(x, s) = 0$. From the observation above, we know that
$$P(\lambda_1, \ldots, \lambda_m)(f_1)_+^{\lambda_1} \cdots (f_m)_+^{\lambda_m} = a(x, \lambda)(f_1)_+^{\lambda_1 - k} \cdots (f_m)_+^{\lambda_m - k}$$
holds if Re $\lambda_i > k$ ($i = 1, \ldots, m$). Since the distribution on the right-hand side is a locally integrable function, it vanishes if and only if $a(x, \lambda) = 0$. The conclusion follows from the uniqueness of analytic continuation. □

**Proposition 2** If the distribution $(f_1)_+^{s_1} \cdots (f_m)_+^{s_m}$ is well-defined and holomorphic in $s = (s_1, \ldots, s_m)$ on a neighborhood of $\lambda = (\lambda_1, \ldots, \lambda_m) \in \mathbb{C}^m$, then there is a surjective $D_n$-homomorphism
$$\mathcal{N}(\lambda) \longrightarrow D_n(f_1)_+^{\lambda_1} \cdots (f_m)_+^{\lambda_m}$$
which sends the residue class $[f_1^{s_1} \cdots f_m^{s_m}] \in \mathcal{N}(\lambda)$ to $(f_1)_+^{\lambda_1} \cdots (f_m)_+^{\lambda_m}$.

Proof: Let us define a surjective $D_n$-homomorphism $\Phi : \mathcal{N}(\lambda) \to D_n(f_1)_+^{\lambda_1} \cdots (f_m)_+^{\lambda_m}$ by
$$\Phi(P[f_1^{s_1} \cdots f_m^{s_m}]) = P(f_1)_+^{\lambda_1} \cdots (f_m)_+^{\lambda_m}$$
for $P \in D_n$. If $P[f_1^{s_1} \cdots f_m^{s_m}] = 0$ in $\mathcal{N}(\lambda)$, then there exist $P_1(s), \ldots, P_m(s) \in D_n[s]$ and $Q[s] \in \text{Ann}_{D_n[s]} f_1^{s_1} \cdots f_m^{s_m}$ such that
$$P = (s_1 - \lambda_1)P_1(s) + \cdots + (s_m - \lambda_m)P_m(s) + Q(s).$$
Then in view of Lemma 1 and the uniqueness of analytic continuation, we have
$$P(f_1)_+^{\lambda_1} \cdots (f_m)_+^{\lambda_m} = Q(\lambda)(f_1)_+^{\lambda_1} \cdots (f_m)_+^{\lambda_m} = 0.$$
Hence $\Phi$ is well-defined as a homomorphism of left $D_n$-modules. □

It is known that there exist a non-zero polynomial $b(s)$ in $s = (s_1, \ldots, s_m)$ and an operator $P(s) \in D_n[s]$ such that
$$P(s)f_1^{s_1+1} \cdots f_m^{s_m+1} = b(s)f_1^{s_1} \cdots f_m^{s_m}, \quad b(s) = \prod_{i=1}^{\nu}(c_{i1}s_1 + \cdots + c_{im}s_m + c_i) \quad (4)$$

with positive integers $c_{ij}$ and positive rational numbers $c_i$. This was proved by Kashiwara [9] for $m = 1$ and by Gyoja [8] for general $m$. An algorithm to compute the ideal consisting of such $b(s)$, which is called the Bernstein-Sato ideal, was given in [18]. For the case $m = 1$, an algorithm to compute such a functional equation was given in [14]. See also [16] for an algorithm to compute a minimal one in the sense that the differential operator on the left-hand side has minimal degree with respect to $\partial_x$ and $s$.

In view of Lemma 1, we obtain from (4)

$$b(\lambda_1,\ldots,\lambda_m)\langle (f_1)_+^{\lambda_1}\cdots(f_m)_+^{\lambda_m},\varphi\rangle = \langle P(\lambda_1,\ldots,\lambda_m)(f_1)_+^{\lambda_1+1}\cdots(f_m)_+^{\lambda_m+1},\varphi\rangle$$
$$= \langle (f_1)_+^{\lambda_1+1}\cdots(f_m)_+^{\lambda_m+1}, {}^tP(\lambda_1,\ldots,\lambda_m)\varphi\rangle$$

for $\varphi \in C_0^\infty(\mathbb{R}^n)$ with the formal adjoint operator ${}^tP(s) \in D_n[s]$ of $P(s)$. By using the functional equation (4) repeatedly, we know that $(f_1)_+^{\lambda_1}\cdots(f_m)_+^{\lambda_m}$ can be continued to a distribution in $x$ which is meromorphic in $(\lambda_1,\ldots,\lambda_m)$ on the whole $\mathbb{C}^m$. More precisely, $(f_1)_+^{\lambda_1}\cdots(f_m)_+^{\lambda_m}$ is holomorphic in $(\lambda_1,\ldots,\lambda_m)$ on

$$\Omega(f_1,\ldots,f_m) := \{(\lambda_1,\ldots,\lambda_m) \in \mathbb{C}^m \mid b(\lambda_1+k,\ldots,\lambda_m+k) \neq 0 \text{ for any } k \in \mathbb{N}\}.$$

**Lemma 2** *The set $\Omega(f_1,\ldots,f_m)$ contains the closure of $\Omega_+$.*

Proof: Assume that $\lambda^{(0)} = (\lambda_1^{(0)},\ldots,\lambda_m^{(0)})$ belongs to the closure of $\Omega_+$, i.e, $\mathrm{Re}\,\lambda_i^{(0)} \geq 0$ for any $i$. Then we have

$$\mathrm{Re}\left(c_{i1}(\lambda_1^{(0)} + k) + \cdots + c_{im}(\lambda_m^{(0)} + k) + c_i\right) \geq c_i > 0$$

for any $k \in \mathbb{N}$. This implies $b(\lambda_1^{(0)} + k,\ldots,\lambda_m^{(0)} + k) \neq 0$. Hence $\lambda^{(0)}$ belongs to $\Omega(f_1,\ldots,f_m)$. $\square$

Propositions 1 and 2 provide us with an algorithm to compute a holonomic system for $(f_1)_+^{\lambda_1}\cdots(f_m)_+^{\lambda_m}$, and hence for $Y(f_1)\cdots Y(f_m)$ by virtue of Lemma 2.

**Algorithm 1 (a holonomic system for $(f_1)_+^{\lambda_1}\cdots(f_m)_+^{\lambda_m}$)**

**Input:** $f_1,\ldots,f_m \in \mathbb{R}[x]$ and $\lambda = (\lambda_1,\ldots,\lambda_m) \in \Omega(f_1,\ldots,f_m)$.
**Output:** A set $G$ of generators of a holonomic ideal contained in $\mathrm{Ann}_{D_n}(f_1)_+^{\lambda_1}\cdots(f_m)_+^{\lambda_m}$.
1. Compute a set $G_1$ of generators of the annihilator $\mathrm{Ann}_{D_n[s]} f_1^{s_1}\cdots f_m^{s_m}$ by an algorithm of [18], or of [4]. Following the former, let $J$ be the left ideal of the ring of differential operators on the variables $x_1,\ldots,x_n,t_1,\ldots,t_m$ which is generated by

$$\partial_{x_i} - \sum_{j=1}^m \frac{\partial f_j}{\partial x_i}\partial_{t_j} \quad (i = 1,\ldots,n), \quad t_j - f_j \quad (j = 1,\ldots,m).$$

Then compute the intersection $J \cap D_n[s]$ with the identification $s_j = -\partial_{t_j}t_j$ ($j = 1,\ldots,m$) by using Algorithm 7 in Section 5.
2. Set $G := \{P(\lambda) \mid P(s) \in G_1\}$.

We do not know if the output of this algorithm coincides with the annihilator ideal of $(f_1)_+^{\lambda_1}\cdots(f_m)_+^{\lambda_m}$. As a trivial counter example, take $f = x^2$ with a variable $x$. Then $\partial_x Y(f) = \partial_x 1 = 0$ holds. On the other hand, the annihilating ideal of $f^s = x^{2s}$ is generated by $x\partial_x - 2s$. Hence $\partial_x$ cannot be obtained by substitution $s = 0$ from an annihilator of $f^s$.

**Example 3** Set $f = x^3 - y^2$. From the functional equation

$$\left(\frac{1}{27}\partial_x^3 + \frac{1}{8}y\partial_y^3 - \frac{4\lambda+3}{8}\partial_y^2\right)f_+^{\lambda+1} = (\lambda+1)\left(\lambda+\frac{5}{6}\right)\left(\lambda+\frac{7}{6}\right)f_+^\lambda,$$

we know that the distribution $u := f_+^\lambda$ is holomorphic in $\lambda$ belonging to the set

$$\Omega(f) = \{\lambda \in \mathbb{C} \mid \lambda \neq -1-k,\ -\frac{5}{6}-k,\ -\frac{7}{6}-k \quad (k=0,1,2,\ldots)\}.$$

A holonomic system for $u$ is given by

$$(2y\partial_x + 3x^2\partial_y)u = (2x\partial_x + 3y\partial_y - 6\lambda)u = 0.$$

In particular, the Heaviside function $Y(f)$ satisfies a holonomic system

$$(2y\partial_x + 3x^2\partial_y)Y(f) = (2x\partial_x + 3y\partial_y)Y(f) = 0.$$

Differentiation of the distribution $(f_1)_+^{\lambda_1}\cdots(f_m)_+^{\lambda_m}$ with respect to the complex parameters $\lambda_1, \ldots, \lambda_m$ yields a distribution

$$(f_1)_+^{\lambda_1}\cdots(f_m)_+^{\lambda_m}(\log f_1)^{\nu_1}\cdots(\log f_m)^{\nu_m},$$

which is holomorphic in $\lambda$ on $\Omega(f_1, \ldots, f_m)$ for arbitrary non-negative integers $\nu_1, \ldots, \nu_m$. A holonomic system for this distribution can be computed through differentiation of the annihilator of $f_1^{s_1}\cdots f_m^{s_m}$ with respect to the parameters $s_1, \ldots, s_m$.

For the sake of simplicity, let us describe an algorithm in case $m = 1$. The general case $m \geq 2$ is quite similar.

**Algorithm 2 (a holonomic system for $f_+^\lambda(\log f)^k$)**

**Input:** $f \in \mathbb{R}[x]$, $k \in \mathbb{N}$ and $\lambda \in \Omega(f)$.
**Output:** a set $G$ of generators of a holonomic ideal contained in $\mathrm{Ann}_{D_n} f^\lambda(\log f)^k$.
1. Compute a set $G_1$ of generators of the annihilator $\mathrm{Ann}_{D_n[s]} f^s$.
2. Let $e_1 = (1,0,\ldots,0), \cdots, e_{k+1} = (0,\ldots,0,1)$ be the canonical base of $\mathbb{Z}^{k+1}$. For each $P(s) \in G_1$ and an integer $j$ with $0 \leq j \leq k$, set

$$P^{(j)}(s) := \sum_{i=0}^{j} \binom{j}{i}\frac{\partial^{j-i}P(s)}{\partial s^{j-i}}e_{i+1} \in (D_n[s])^{k+1}.$$

3. Let $N$ be the submodule of $D_n[s]^{k+1}$ generated by the set $\{P^{(j)}(s) \mid P(s) \in G_1,\ 0 \leq j \leq k\}$. Compute a set $G_2$ of generators of the ideal

$$I := \{Q(s) \in D_n[s] \mid Q(s)e_{k+1} \in N\}$$

by eliminating $e_1, \ldots, e_k$ via a Gröbner base of $N$ with respect to a 'position over term' ordering.
4. Set $G := \{Q(\lambda) \mid Q(s) \in G_2\}$.

**Proposition 3** *The ideal $I$ of the preceding algorithm is holonomic and annihilates the distribution $f_+^\lambda (\log f)^k$ for any $\lambda \in \Omega(f)$.*

Proof: Let $P(s)$ belong to $G_1$. Differentiating the equation $P(s)f^s = 0$ with respect to $s$, one gets

$$\sum_{i=0}^{j} \binom{j}{i} \frac{\partial^{j-i} P(s)}{\partial s^{j-i}} f^s (\log f)^i = 0.$$

Hence $P^{(j)}(\lambda)$ annihilates the vector $(f_+^\lambda, f_+^\lambda \log f, \ldots, f_+^\lambda (\log f)^k)$ of distributions for $\lambda \in \Omega(f)$. It follows that each element of $G$ annihilates $f_+^\lambda (\log f)^k$.

In order to prove the holonomicity, let $N(\lambda)$ be the submodule of $(D_n)^{k+1}$ generated by $\{P(\lambda) \mid P(s) \in N\}$. Then it suffices to prove that $(D_n)^{k+1}/N(\lambda)$ is holonomic since $D_n/I$ is its submodule. We regard $(D_n)^{j+1}$ as the submodule of $(D_n)^{k+1}$ generated by $e_1, \ldots, e_{j+1}$ if $0 \leq j \leq k$. Set

$$\mathcal{N}(\lambda, j) := (D_n)^{j+1}/(N(\lambda) \cap (D_n)^{j+1}).$$

Then we have an increasing sequence

$$\mathcal{N}(\lambda, 0) \subset \mathcal{N}(\lambda, 1) \subset \cdots \subset \mathcal{N}(\lambda, k) = (D_n)^{k+1}/N(\lambda)$$

of left $D_n$-modules. We have only to show that $\mathcal{N}(\lambda, 0)$ and $\mathcal{N}(\lambda, j)/\mathcal{N}(\lambda, j-1)$ are holonomic for $1 \leq j \leq k$.

Note that $\mathcal{N}(\lambda)$, which annihilates $f_+^\lambda$, is isomorphic to $D_n/I(\lambda)$ with

$$I(\lambda) := \{P(\lambda) \mid P(s) \in \mathrm{Ann}_{D_n[s]} f^s\}.$$

From the definition we have $I(\lambda) \subset N(\lambda) \cap D_n$. Together with Proposition 1 this implies that $\mathcal{N}(\lambda, 0)$ is holonomic. Let $\iota_j$ be the homomorphism of $D_n$ to $(D_n)^{j+1}$ which sends $P \in D_n$ to $Pe_{j+1}$. It is easy to see that $I(\lambda)e_{j+1}$ is contained in $(N(\lambda) \cap (D_n)^{j+1}) + (D_n)^j$. Hence $\iota_j$ induces a surjective homomorphism of the holonomic module $\mathcal{N}(\lambda)$ to $\mathcal{N}(\lambda, j)/\mathcal{N}(\lambda, j-1)$, which is hence also holonomic. This completes the proof. □

**Example 4** Set $f = x^3 - y^2$. Then Algorithm 2 outputs three annihilators

$2y\partial_x + 3x^2 \partial_y,$
$4x^2 \partial_x^2 + (12yx\partial_y + (-24\lambda + 4)x)\partial_x + 9y^2 \partial_y^2 + (-36\lambda + 9)y\partial_y + 36\lambda^2,$
$-8y\partial_x^3 + (36yx\partial_y^2 - 72\lambda x \partial_y)\partial_x + 27y^2 \partial_y^3 + (-108\lambda + 81)y\partial_y^2 + (108\lambda^2 - 108\lambda + 3)\partial_y$

of $f_+^\lambda \log f$.

Finally let us consider the delta function $\delta(f)$ for a non-singular real polynomial $f \in \mathbb{R}[x]$ in $n$ variables. As distribution, $\delta(f)$ is defined to be the integral

$$\int_{f=0} \varphi \omega$$

for $\varphi \in C_0^\infty(\mathbb{R}^n)$, where $\omega$ is the volume element of the hypersurface $f = 0$ such that $df \wedge \omega = dx_1 \wedge \cdots \wedge dx_n$. The following should be well-known:

**Proposition 4** *Let $f \in \mathbb{R}[x]$ be non-singular in $\mathbb{C}^n$; i.e., the variety defined by*

$$\{x \in \mathbb{C}^n \mid f(x) = \frac{\partial f}{\partial x_1}(x) = \cdots = \frac{\partial f}{\partial x_n}(x) = 0\}$$

*be empty. Then $\delta(f)$ satisfies a holonomic system*

$$f\delta(f) = 0, \quad \left(\frac{\partial f}{\partial x_j}\partial_i - \frac{\partial f}{\partial x_i}\partial_j\right)\delta(f) = 0 \quad (1 \le i < j \le n).$$

Proof: The characteristic variety is contained in the $n$-dimensional algebraic set

$$\left\{(x,\xi) \in \mathbb{C}^{2n} \mid f(x) = 0, \ \xi_i = c\frac{\partial f}{\partial x_i} \ (i=1,\ldots,n), \ c \in \mathbb{C}\right\}.$$

Hence the system above is holonomic. □

## 3 Operations on holonomic functions

Let us assume that $u$ and $v$ are holonomic functions or distributions. Then the following functions or distributions are holonomic under the condition that they are 'well-defined'. Moreover, if a holonomic ideal for $u$ (and also one for $v$ if relevant) is explicitly given, then there exist algorithms to compute a holonomic ideal which each of the following functions satisfies.

1. $Pu$ with $P \in D_n$.

2. The sum $u + v$.

3. The restriction $u|_Y$ of $u$ to an affine subspace $Y$ of $\mathbb{R}^n$ if it is well-defined as in the case where $u$ is smooth.

4. The product $uv$ if it is well-defined as in the case where $u$ is $C^\infty$ and $v$ is a distribution.

5. The definite integral $\displaystyle\int_{\mathbb{R}^d} u(x_1,\ldots,x_{n-d},x_{n-d+1},\ldots,x_n)\,dx_{n-d+1}\cdots dx_n$ with parameters if it is well-defined as in the case where $u$ has a compact support with respect to the integration variables.

Let us explain briefly how to compute holonomic ideals for functions above. The algorithms which will be presented in this section are more or less well-known at present. Nevertheless, we would like to recall them in order to verify that they certainly apply to holonomic distributions as well as for the reader's convenience.

Let $I$ and $J$ be holonomic ideals for $u$ and $v$ respectively. First, a holonomic ideal for $Pu$ can be computed as an ideal quotient $I : P$ by using Gröbner bases in the same way as in the polynomial ring. The left $D_n$-homomorphism of $D_n$ to itself which sends $Q \in D_n$ to $QP$ induces an injective homomorphism $D_n/(I:P) \to D_n/I$. Hence $D_n/(I:P)$ is holonomic.

Second, a holonomic ideal for $u+v$ can be computed as an ideal intersection $I \cap J$. The left $D_n$-homomorphism of $D_n$ to $(D_n)^2$ which sends $Q \in D_n$ to $(Q, -Q)$ induces an injective homomorphism

$$D_n/(I \cap J) \to (D/I) \oplus (D/J).$$

This implies that $D_n/(I \cap J)$ is holonomic.

The restriction algorithm was given in [15] for one codimensional case and in [19] for the general case. The restriction algorithm is translated to the integration algorithm through the algebraic Fourier transformation of the ring of differential operators. Hence let us describe the integration algorithm instead.

Let $u(x,t)$ be a holonomic distribution defined on $U \times \mathbb{R}^d$ in the variables $(x,t) = (x_1, \ldots, x_n, t_1, \ldots, t_d)$, where $U$ is an open set of $\mathbb{R}^n$. Then the definite integral of $u(x,t)$ with respect to $t = (t_1, \ldots, t_d)$ is defined as the distribution

$$C_0^\infty(U) \ni \varphi(x) \longmapsto \langle \int_{\mathbb{R}^d} u(x,t)\, dt_1 \cdots dt_d,\, \varphi \rangle = \langle u(x,t), \varphi(x)1(t) \rangle,$$

where $1(t)$ denotes the identity function which takes only the value 1 for all $t$. This definite integral is well-defined if $u(x,t)$ is a distribution with proper support with respect to $t$, i.e., for each $x_0 \in U$, there exists a neighborhood $V$ of $x_0$ and a compact set $K$ of $\mathbb{R}^d$ such that $u(x,t)$ vanishes on $V \times (\mathbb{R}^d \setminus K)$, or else if $u(x,t)$ is $C^\infty$ in $(x,t)$ and rapidly decreasing with respect to $t$, i.e., for any $x_0$ and $\alpha, \beta \in \mathbb{N}^d$, there exists a neighborhood $V$ of $x_0$ such that $t^\alpha \partial_t^\beta u(x,t)$ is bounded on $V \times \mathbb{R}^d$.

Let $I$ be a left ideal of $D_{n+d}$ which is contained in $\operatorname{Ann}_{D_{n+d}} u(x,t)$. The $D$-module theoretic integration of the module $D_{n+d}/I$ is defined to be the left $D_n$-module

$$\int D_{n+d}/I\, dt := D_{n+d}/(\partial_{t_1} D_{n+d} + \cdots + \partial_{t_d} D_{n+d} + I).$$

Let $w = (w_1, \ldots, w_{2n+2d}) \in \mathbb{Z}^{2(n+d)}$ be the weight vector for $D_{n+d}$ such that $w_{n+i} = 1$ (the weight for $t_i$) and $w_{2n+d+i} = -1$ (the weight for $\partial_{t_i}$) for $i = 1, \ldots, d$ and that other components of $w$ are zero. If $I$ is a holonomic ideal, then there exists a nonzero univariate polynomial $b(s)$ (the $b$-function of $I$ with respect to $w$) of the minimum degree such that

$$b(-\partial_{t_1} t_1 - \cdots - \partial_{t_d} t_d) + P \in I$$

with some $P \in F_{-1}^w(D_{n+d})$. Let $k_1$ be the maximum integral root of $b(s) = 0$. (Set $k_1 = -1$ if there is none.) Then the $D_n$-module $\int D_{n+d}/I\, dt$ is generated by the residue classes of $t^\alpha$ with $\alpha \in \mathbb{N}^d$ such that $|\alpha| \leq k_1$. Set

$$N := \{(P_\alpha)_\alpha \in \bigoplus_{|\alpha| \leq k_1} D_n \mid \sum_{|\alpha| \leq k_1} P_\alpha t^\alpha \in \partial_{t_1} D_{n+d} + \cdots + \partial_{t_d} D_{n+d} + I\}.$$

The $b$-function $b(s)$ and a set of generators of $N$ can be computed by a Gröbner base of $I$ with respect to the weight vector $w$. Then we have an isomorphism $\int D_{n+d}/I\, dt \simeq \left( \bigoplus_{|\alpha| \leq k_1} D_n \right) / N$, which is a holonomic $D_n$-module.

**Definition 4** The *integration ideal* of a left ideal $I$ of $D_{n+d}$ is defined to be the left ideal
$$N_0 := \{P \in D_n \mid P \in \partial_{t_1} D_{n+d} + \cdots + \partial_{t_d} D_{n+d} + I\}$$
of $D_n$.

Once a set of generators of $N$ is obtained, the integration ideal $N_0$ can be computed by elimination in the free module $\oplus_{|\alpha| \leq k_1} D_n$.

**Theorem 1** *Suppose that the definite integral $v(x) := \int_{\mathbb{R}^d} u(x,t)\, dt_1 \cdots dt_d$ is well-defined as a distribution on an open set $U$ of $\mathbb{R}^n$ and $I$ is a holonomic ideal of $D_{n+d}$ annihilating $u(x,t)$. Then the integration ideal $N_0$ of $I$ is holonomic and annihilates $v(x)$.*

Proof: Let $P = P(x, \partial_x) \in D_n$ belong to $N_0$. Then there exist $Q_i = Q_i(x,t,\partial_x,\partial_t) \in D_{n+d}$ and $R \in I$ such that

$$P = \partial_{t_1} Q_1 + \cdots + \partial_{t_d} Q_d + R.$$

Thus for any $\varphi(x) \in C_0^\infty(U)$ one has

$$\langle P \int_{\mathbb{R}^d} u(x,t)\, dt_1 \cdots dt_d,\ \varphi \rangle = \langle Pu(x,t),\ \varphi(x) 1(t) \rangle$$

$$= \sum_{i=1}^d \langle \partial_{t_i} Q_i u(x,t),\ \varphi(x) 1(t) \rangle$$

$$= -\sum_{i=1}^d \langle u(x,t),\ {}^t Q_i \partial_{t_i}(\varphi(x) 1(t)) \rangle = 0,$$

where ${}^t P$ denotes the formal adjoint of $P$.

The $D_n$-module $D_n/N_0$ is holonomic since it can be regarded as a submodule of the holonomic module $\bigoplus_{|\alpha| \leq k_1} D_n/N$, the proof of the holonomicity of which is given, e.g, in [3]. □

The integration algorithm is summarized as follows. See [18] or [21] for the correctness proof.

**Algorithm 3 (integration ideal)**

**Input:** A set $G_0$ of generators of a holonomic ideal $I$ of $D_{n+d}$ annihilating $u(x,t)$.
**Output:** A set $G$ of generators of a holonomic ideal $N_0$ of $D_n$ annihilating $v(x) := \int_{\mathbb{R}^d} u(x,t)\, dt_1 \cdots dt_d$.
1. Compute a Gröbner base $G_1$ of $I$ with respect to a monomial order which is compatible with the weight vector $w = (0,\ldots,0,1,\ldots,1; 0,\ldots,0,-1,\ldots,-1)$ for the variables $(x,t,\partial_x,\partial_t)$.
2. Compute a $b$-function of $I$ with respect to $w$, which is a univariate polynomial $b(s)$ of the minimum degree such that $b(-\partial_{t_1} t_1 - \cdots - \partial_{t_d} t_d) + P$ belongs to $I$ with some $P \in F_{-1}^w(D_{n+d})$. This amounts to computing a generator of the intersection ideal

$$\mathbb{C}[\partial_{t_1} t_1 + \cdots + \partial_{t_d} t_d] \cap \bigoplus_{k \in \mathbb{Z}} F_k^w(I)/F_{k-1}^w(I).$$

3. Let $k_1$ be the maximum integral root of $b(s) = 0$ if any; if there is none or else $k_1 < 0$, then set $G := \{1\}$ and quit.
4. For $P \in G_1$ and $\alpha \in \mathbb{N}^d$ such that $\mathrm{ord}_w(P) + |\alpha| \leq k_1$, one has an expression of the form

$$t^\alpha P = \sum_{j=1}^d \partial_{t_j} Q_j + \sum_{|\beta| \leq k_1} R_\beta t^\beta$$

with $Q_j \in D_{n+d}$ and $R_\beta \in D_n$. Set $\chi(t^\alpha P) := (R_\beta)_{|\beta| \leq k_1}$. Let $N$ be the submodule of $\bigoplus_{|\beta| \leq k_1} D_n$ generated by

$$\{\chi(t^\alpha P) \mid P \in G_1,\ \mathrm{ord}_w(P) + |\alpha| \leq k_1\}.$$

5. Compute a set $G$ of generators of $N \cap D_n$ via a Gröbner base of $N$ with an appropriate 'position over term' ordering, where $D_n$ is identified with the submodule $\bigoplus_{|\beta| \leq 0} D_n$ of $\bigoplus_{|\beta| \leq k_1} D_n$.

Finally let us describe an algorithm for the product. Let $u(x)$ and $v(x)$ be holonomic distributions defined on $\mathbb{R}^n$ and assume that the product $(Pu(x))(Qv(x))$ is well-defined as a distribution for any operators $P, Q \in D_n$ and satisfies the following 'associativity' condition:

$$(a(x)Pu)(Qv) = (Pu)(a(x)Qv) \qquad (\forall P, Q \in D_n,\ \forall a(x) \in \mathbb{C}[x]). \tag{5}$$

This is certainly the case, for example, when either $u(x)$ or $v(x)$ is $C^\infty$.

Let $I$ and $J$ be left ideals of $D_n$ such that

$$I \subset \mathrm{Ann}_{D_n} u, \quad J \subset \mathrm{Ann}_{D_n} v$$

and set $M := D_n/I$ and $N := D_n/J$. Let $\tilde{u}$ and $\tilde{v}$ be the modulo classes of $1 \in D_n$ in $M$ and $N$ respectively. Then there is a natural bilinear map

$$\Psi : M \times N \ni (P\tilde{u}, Q\tilde{v}) \longmapsto (Pu)(Qv) \in D_n(uv) \subset \mathcal{D}'(\mathbb{R}^n)$$

with the property $\Psi(aP\tilde{u}, Q\tilde{v}) = \Psi(P\tilde{u}, aQ\tilde{v})$ for any $a \in \mathbb{C}[x]$. Hence by the universal property of tensor product, there is a $D_n$-homomorphism

$$\Phi : M \otimes_{\mathbb{C}[x]} N \to D_n(uv)$$

such that $\Phi((P\tilde{u}) \otimes (Q\tilde{v})) = \Psi(P\tilde{u}, Q\tilde{v}) = (Pu)(Qv)$ for any $P, Q \in D_n$. Moreover, $M \otimes_{\mathbb{C}[x]} N$ has a natural structure of left $D_n$-module and $\Phi$ gives a homomorphism of $D_n$-module. This means that $P(\tilde{u} \otimes \tilde{v}) = 0$ in $M \otimes N$ implies $P(uv) = 0$ in $\mathcal{D}'(\mathbb{R}^n)$. It is also well-known that $M \otimes_{\mathbb{C}[x]} N$ is holonomic if $M$ and $N$ are holonomic. (See e.g., [3]). In conclusion we have

**Proposition 5** *Let $u, v$ be distributions which satisfy (5). Let $I$ and $J$ be holonomic left ideals of $D_n$ which annihilate $u$ and $v$ respectively. Then the left ideal*

$$\mathrm{Ann}_{D_n}(\tilde{u} \otimes \tilde{v}) := \{P \in D_n \mid P(\tilde{u} \otimes \tilde{v}) = 0\ \text{in}\ M \otimes_{\mathbb{C}[x]} N\}$$

*is holonomic and annihilates the distribution $uv$.*

The 'algebraic' annihilator $\mathrm{Ann}_{D_n}(\tilde{u} \otimes \tilde{v})$ can be computed as follows: First let $I \hat{\otimes} J$ be the left ideal of $D_{2n}$ which are generated by $\{P(x, \partial_x) \mid P \in I\}$ and $\{Q(y, \partial_y) \mid Q \in J\}$. Here we use the notation $\partial_x = (\partial/\partial x_1, \ldots, \partial/\partial x_n)$ and $\partial_y = (\partial/\partial y_1, \ldots, \partial/\partial y_n)$. We change the variables $(x, y)$ to $(x, z)$ by the substitution $x = x,\ z = y - x$. Then the derivations are transformed by

$$\partial_{x_i} = \partial_{x_i} - \partial_{z_i}, \quad \partial_{y_i} = \partial_{z_i} \quad (i = 1, \ldots, n).$$

We regard $I\hat{\otimes}J$ as a left ideal of the ring of differential operators $D_{x,z}$ with respect to the variables $(x,z)$. Then the restriction of $D_{x,z}/I\hat{\otimes}J$ to $z = 0$ is defined to be
$$D_{x,z}/(I\hat{\otimes}J + z_1 D_{x,z} + \cdots + z_n D_{x,z})$$
as left $D_n$-module, where $D_n$ stands for the ring of differential operators with respect to the variables $x = (x_1, \ldots, x_n)$. One has an isomorphism
$$M \otimes_{\mathbb{C}[x]} N \simeq D_{x,z}/(I\hat{\otimes}J + z_1 D_{x,z} + \cdots + z_n D_{x,z}).$$

Set
$$I\hat{\otimes}J|_\Delta := \{P(x,\partial_x) \in D_n \mid P(x,\partial_x) - \sum_{i=1}^n z_i Q_i \in I\hat{\otimes}J \quad (\exists Q_1, \ldots, Q_n \in D_{x,z})\}.$$

This ideal can be computed by the restriction algorithm and coincides with $\mathrm{Ann}_{D_n}(\tilde{u} \otimes \tilde{v})$ in $M \otimes_{\mathbb{C}[x]} N$. See [19] for details.

For practical computations of restriction, integration and tensor product, we have made use of a library file `nk_restriction.rr` written by H. Nakayama for a computer algebra system Risa/Asir [12].

**Example 5** Let us consider the integral
$$v(t) := \int_{\mathbb{R}^3} f e^g \delta(t - x^2 - y^2 - z^2)\, dxdydz = Y(t) \int_{S^2_t} f e^g \omega_t$$

for polynomials $f, g$, where $\omega_t$ denotes the canonical volume element of the sphere $S^2_t := \{(x,y,z) \in \mathbb{R}^3 \mid x^2 + y^2 + z^2 = t\}$. The annihilator ideal $I$ of $\delta(t - x^2 - y^2 - z^2)$ is generated by $t - x^2 - y^2 - z^2$, $\partial_x + 2x\partial_t$, $\partial_y + 2y\partial_t$, $\partial_z + 2z\partial_t$. Then the annihilator of $f\delta(t - x^2 - y^2 - z^2)$ is given by the ideal quotient $I : f$. Finally, the annihilator of $f e^g \delta(t - x^2 - y^2 - z^2)$ is given by the ideal $e^g(I : f)e^{-g}$. For example, if $f = 1$ and $g = x - y^2 - z^2$, we get a differential equation
$$(4t\partial_t^3 + (4t+6)\partial_t^2 + 5\partial_t - 1)v(t) = 0,$$
which has a regular singularity at $t = 0$ with exponents $0, \frac{1}{2}, 1$.

On the other hand, if $f = 2x^2 - y^2 - z^2$ and $g = 0$, we get $v(t) = 0$.

**Example 6** The integral
$$v(t) := \int_{\mathbb{R}^2} \delta(t - x^3 + y^2) e^{-x^2 - y^2}\, dxdy$$

satisfies a differential equation
$$(108t^2\partial_t^5 + (-216t^2 + 648t)\partial_t^4 + (108t^2 - 972t + 627)\partial_t^3$$
$$+ (356t - 606)\partial_t^2 + (-64t + 108)\partial_t + 32t - 48)v(t) = 0,$$

which has a regular singularity at $t = 0$ with exponents $0, 1, 2, -\frac{1}{6}, \frac{7}{6}$.

# 4 Definite integrals with the Heaviside function

Let $u$ be a holonomic function on $\mathbb{R}^n$. Let $f_1, \ldots, f_m$ be nonzero polynomials in $x = (x_1, \ldots, x_n)$ with real coefficients. Setting

$$D(x_1, \ldots, x_{n-d}) = \{(x_{n-d+1}, \ldots, x_n) \in \mathbb{R}^d \mid f_1(x) \geq 0, \ldots, f_m(x) \geq 0\},$$

let us consider the definite integral

$$v(x_1, \ldots, x_{n-d}) = \int_{D(x_1, \ldots, x_{n-d})} u(x)\, dx_{n-d+1} \cdots dx_n$$

$$= \int_{\mathbb{R}^n} Y(f_1) \cdots Y(f_m) u(x)\, dx_{n-d+1} \cdots dx_n.$$

We suppose that the set $\{f_j(x) > 0 \mid j = 1, \ldots, m\}$ is non-empty and the integral is well-defined if $(x_1, \ldots, x_{n-d})$ belongs to an open set $U$ of $\mathbb{R}^{n-d}$. This is the case, for example, when $u$ is a $C^\infty$ function on an open set $W$ of $\mathbb{R}^n$ such that

$$\{(x_1, \ldots, x_{n-d})\} \times D(x_1, \ldots, x_{n-d}) \subset W$$

and $D(x_1, \ldots, x_{n-d})$ is compact if $(x_1, \ldots, x_{n-d}) \in U$. Then a holonomic ideal for this integral can be computed by combining the algorithms explained in the preceding two sections. In conclusion, we have the following algorithm.

**Algorithm 4 (a holonomic ideal for a definite integral)**

**Input:** a holonomic ideal $J$ annihilating the function $u$.
**Output:** a holonomic ideal $J$ annihilating the definite integral $v$.
1. Compute a holonomic ideal annihilating $Y(f_1) \cdots Y(f_m)$ by Algorithm 1.
2. Compute a holonomic ideal $I$ annihilating the product $uY(f_1) \cdots Y(f_m)$ by using the tensor product computation.
3. Compute the integration ideal of $I$ with respect to $x_{n-d+1}, \ldots, x_n$ by Algorithm 3.

In the step 2 of this algorithm, if $u$ is of the form $u = (g_1)_+^{\lambda_1} \cdots (g_p)_+^{\lambda_p} e^h$ with polynomials $g_1, \ldots, g_p, h$, then we can compute first a set $G$ which generates a holonomic annihilating ideal $I$ of

$$(f_1)_+^0 \cdots (f_m)_+^0 (g_1)_+^{\lambda_1} \cdots (g_p)_+^{\lambda_p}$$

by using Algorithm 1. Then an annihilating ideal for $uY(f_1) \cdots Y(f_p)$ can be computed as an ideal $e^h I e^{-h}$ of $D_n$. A set of generators of this ideal is obtained by substituting $\partial_i - \partial h/\partial_i$ for $\partial_i$ ($i = 1, \ldots, n$) in each element of $G$.

**Example 7** Set

$$v(t) = \int_{x^2+y^2 \leq t} \frac{dxdy}{1 + x^4 + y^4} = \int_{\mathbb{R}^2} Y(t - x^2 - y^2)(1 + x^4 + y^4)^{-1}\, dxdy.$$

First by Algorithm 1, we compute a holonomic system for the integrand

$$u(x, y, t) = Y(t - x^2 - y^2)(1 + x^4 + y^4)^{-1}.$$

Then the integration algorithm outputs a holonomic system

$$((t^5 + 3t^3 + 2t)\partial_t^2 + (2t^4 + 3t^2)\partial_t)v(t) = 0.$$

Solving this differential equation by quadratures, one gets

$$v(t) = C_1 \int_0^t \frac{ds}{\sqrt{s^4 + 3s^2 + 2}} + C_2$$

for $t > 0$ with some constants $C_1, C_2$. On the other hand, by the change of variables to polar coordinates, one has

$$v(0) = 0, \quad \lim_{t \to +0} v'(t) = \pi.$$

In conclusion, we have proved the identity

$$v(t) = \sqrt{2}\pi Y(t) \int_0^t \frac{ds}{\sqrt{s^4 + 3s^2 + 2}}.$$

**Example 8** Set

$$v(t) = \int_{x^6 + x^4 y^2 + y^4 \leq t} dxdy = \int_{\mathbb{R}^2} Y(t - x^6 - x^4 y^2 - y^4) \, dxdy.$$

By integrating a holonomic system for the integrand, we obtain a differential equation

$$((147456t^7 - 995328t^6)\partial_t^7 + (3096576t^6 - 15925248t^5)\partial_t^6$$
$$+ (20604416t^5 - 74822400t^4)\partial_t^5 + (51215360t^4 - 115430400t^3)\partial_t^4$$
$$+ (43401540t^3 - 46770960t^2)\partial_t^3 + (8707020t^2 - 2078400t)\partial_t^2$$
$$+ (110880t - 105)\partial_t)v(t) = 0.$$

This differential equation has a regular singularity at $t = 0$ with exponents $0$, $\frac{5}{12}, \frac{7}{12}, \frac{9}{12}, \frac{11}{12}, \frac{13}{12}, \frac{15}{12}$. It seems difficult to identify $v(t)$ in the 7-dimensional solution space of the above equation.

**Example 9** The integral

$$v(t) = \int_{x^3 - y^2 \geq 0} e^{-t(x^2 + y^2)} \, dxdy = \int_{\mathbb{R}^2} Y(x^3 - y^2) e^{-t(x^2 + y^2)} \, dxdy$$

satisfies a differential equation

$$(216t^4\partial_t^4 + (32t^4 + 1836t^3)\partial_t^3 + (224t^3 + 3594t^2)\partial_t^2 + (326t^2 + 1371t)\partial_t + 70t + 15)v(t) = 0$$

for $t > 0$.

## 5 Difference-differential equations for definite integrals

Let us consider the case where the integrand has some auxiliary parameters $a = (a_1, \ldots, a_p)$ other than $x = (x_1, \cdots, x_n)$. The general algorithm given in the preceding section does not work directly since the integrand does not necessarily satisfies a holonomic system including the parameters $a$ as variables. To avoid this drawback, we consider difference-differential equations including the parameters.

**Definition 5** A subset $\Omega$ of $\mathbb{C}^p$ is said to be *shift-invariant* if $a \in \Omega$ implies that $a + (1, 0, \ldots, 0), \ldots, a + (0, \ldots, 0, 1)$ also belong to $\Omega$.

Let $u(x, a)$ be a distribution in $x = (x_1, \ldots, x_n)$ on an open set $U$ of $\mathbb{R}^n$ with parameters $a = (a_1, \ldots, a_p)$ which belong to a shift-invariant subset $\Omega$ of $\mathbb{C}^p$. Then the shift operator $E_{a_i}$ with respect to $a_i$ acts on $u(x, a)$ by

$$E_{a_i} u(x, a_1, \ldots, a_{i-1}, a_i, a_{i+1}, \ldots, a_p) = u(x, a_1, \ldots, a_{i-1}, a_i+1, a_{i+1}, \ldots, a_p) \quad (i = 1, \ldots, p).$$

Let $D_n$ be the ring of differential operators on the variables $x$. We denote by $D_n \langle a, E_a \rangle$ the ring of difference-differential operators which is generated by $a_i$ and $E_{a_i}$ ($i = 1, \ldots, p$) over $D_n$ with the commutation relations

$$E_{a_i} a_i - a_i E_{a_i} = \begin{cases} E_{a_i} & (i = j) \\ 0 & (i \neq j) \end{cases}, \quad E_{a_i} E_{a_j} = E_{a_j} E_{a_i}, \quad a_i a_j = a_j a_i,$$

where we assume that $a_i$ and $E_{a_i}$ commute with the elements of $D_n$. We introduce new variables $t = (t_1, \ldots, t_p)$ and the associated derivations $\partial_t = (\partial_{t_1}, \ldots, \partial_{t_p})$ and consider the ring $D_{n+p}$ of differential operators on the variables $(x, t) = (x_1, \ldots, x_n, t_1, \ldots, t_p)$. Let $\mu : D_{n+p} \to D_n \langle a, E_a, E_a^{-1} \rangle$ be the homomorphism of $D_n$-algebra defined by

$$\mu(t_i) = E_{a_i}, \quad \mu(\partial_{t_i}) = -a_i E_{a_i}^{-1}.$$

This homomorphism is well-defined since

$$\mu(\partial_{t_i} t_i - t_i \partial_{t_i}) = \mu(\partial_{t_i})\mu(t_i) - \mu(t_i)\mu(\partial_{t_i}) = -a_i E_{a_i}^{-1} E_{a_i} - E_{a_i}(-a_i) E_{a_i}^{-1} = 1.$$

It is easy to see that $\mu$ is injective and can be extended to an isomorphism of $D_{n+p}[(t_1 \ldots t_p)^{-1}]$ to $D_n \langle a, E_a, E_a^{-1} \rangle$.

The inverse shift $E_{a_i}^{-1}$ 'acts' on $u(x, a)$ by

$$E_{a_i}^{-1} u(x, a_1, \ldots, a_{i-1}, a_i, a_{i+1}, \ldots, a_p) = u(x, a_1, \ldots, a_{i-1}, a_i - 1, a_{i+1}, \ldots, a_p)$$

if $(a_1, \ldots, a_i - 1, \ldots, a_p) \in \Omega$. In general, an element $P$ of $D_n \langle a, E_a, E_a^{-1} \rangle$ acts on $u(x, a)$ if $a - k(1, \ldots, 1) \in \Omega$ for a sufficiently large integer $k$. Then an element $Q$ of $D_{n+p}$ acts on $u(x, a)$ by $Qu(x, a) := \mu(Q)u(x, a)$.

In order to justify working in the extended ring $D_n \langle a, E_a, E_a^{-1} \rangle$, or in $D_{n+p}$, let us make the following

**Assumption:** Let $u(x, a)$ be a distribution in $x$ with parameters $a$ defined on $V \times \mathbb{R}^d \times \Omega$ with an open set $V$ of $\mathbb{R}^{n-d}$ and a shift-invariant set $\Omega$ of $\mathbb{C}^p$. The integral

$$v(x', a) := \int_{\mathbb{R}^d} u(x, a)\, dx_{n-d+1} \cdots dx_n$$

with $x' = (x_1, \ldots, x_{n-d})$ is well-defined for $(x', a) \in V \times \Omega$. Let $P$ be an arbitrary element of $D_{n-d} \langle a, E_a, E_a^{-1} \rangle$ such that $Pv(x', a) = 0$ if $x' \in V$ and $a - k(1, \ldots, 1) \in \Omega$ with a sufficiently large $k$. Then for any $\alpha \in \mathbb{N}^p$ such that $E_a^\alpha P \in D_{n-d} \langle a, E_a \rangle$, one has $E_a^\alpha P v(x', a) = 0$ for $x' \in V$ and $a \in \Omega$.

This assumption is satisfied if $v(x', a)$ is holomorphic in $a$ which belongs to a shift-invariant open set $\Omega$ by virtue of the uniqueness of analytic continuation, or else if $\Omega = \mathbb{Z}^p$.

**Definition 6** *A left ideal $I$ of $D_n\langle a, E_a\rangle$ is called a holonomic $D_n\langle a, E_a\rangle$-ideal if*
$$\mu^{-1}(D_n\langle a, E_a, E_a^{-1}\rangle I)$$
*is a holonomic ideal of $D_{n+p}$.*

**Lemma 3** *If a left ideal $J$ of $D_{n+p}$ is holonomic, then $\mu(J) \cap D_n\langle a, E_a\rangle$ is a holonomic $D_n\langle a, E_a\rangle$-ideal.*

Proof: It is sufficient to show the inclusion
$$\mu^{-1}(D_n\langle a, E_a, E_a^{-1}\rangle\,(\mu(J) \cap D_n\langle a, E_a\rangle)) \supset J.$$

If $P$ belongs to $J$, then there exists a positive integer $k$ such that
$$Q := E_{a_1}^k \cdots E_{a_p}^k \mu(P) = \mu(t_1^k \cdots t_p^k P) \in D_n\langle a, E_a\rangle \cap \mu(J).$$

Hence $\mu(P) = E_{s_1}^{-k} \cdots E_{s_p}^{-k} Q$ belongs to $D_n\langle a, E_a, E_a^{-1}\rangle\,(\mu(J) \cap D_n\langle a, E_a\rangle)$. This completes the proof. $\square$

Now let us describe an algorithm to compute a holonomic difference-differential system for the integral $v(x', a)$.

**Algorithm 5 (difference-differential equations for an integral)**

**Input:** A set $G_0$ of generators of a holonomic ideal $J$ of $D_{n+p}$ annihilating $u(x, a)$.
**Output:** A set $G$ of generators of a holonomic ideal of $D_{n-d}\langle a, E_a\rangle$ annihilating the integral $v(x', a)$.
1. Compute a set $G_1$ of generators of the integration ideal
$$N_0 := D_{n-d}\langle t, \partial_t\rangle \cap (\partial_{n-d+1} D_{n+p} + \cdots + \partial_n D_{n+p} + J)$$
of $J$ with respect to $x_{n-d+1}, \ldots, x_n$.
2. Let $P$ be an element of $G_1$. Then there exists a minimal $\nu = (\nu_1, \ldots, \nu_p) \in \mathbb{N}^p$ such that $Q := E_a^\nu \mu(P)$ belongs to $D_{n-d}\langle a, E_a\rangle$. Let us denote this $Q$ by $\mathrm{nm}(\mu(P))$. Set $G := \{\mathrm{nm}(\mu(P)) \mid P \in G_1\}$.

**Theorem 2** *The left ideal of $D_{n-d}\langle a, E_a\rangle$ which is generated by $G$ is holonomic and annihilates $v(x', a)$ for $(x', a) \in V \times \Omega$.*

Proof: If $P$ belongs to $G_1$, then there exist $Q_i \in D_{n+p}$ and $R \in J$ such that
$$P = \sum_{i=n-d+1}^{n} \partial_{x_i} Q_i + R.$$

There exists $\nu \in \mathbb{N}^p$ such that $Q := E_a^\nu \mu(P)$, $E_a^\nu \mu(\partial_{x_i} Q_i)$ and $E_a^\nu \mu(R)$ all belong to $D_n\langle a, E_a\rangle$. Then one has
$$Qv(x', a) = \int_{\mathbb{R}^d} Qu(x, a)\, dx_{n-d+1} \cdots dx_n$$
$$= \sum_{i=n-d+1}^{n} \int_{\mathbb{R}^d} \partial_{x_i} E_a^\nu \mu(Q_i) u(x, a)\, dx_{n-d+1} \cdots dx_n = 0$$

if $a - k(1, \ldots, 1) \in \Omega$ for a sufficiently large $k$. Then it follows from Assumption that $\mu(P)v = 0$ holds for any $a \in \Omega$. The holonomicity of the output follows from Theorem 1 and Lemma 3. □

As a typical example, let $u$ be a holonomic distribution defined on an open set $U$ of $\mathbb{R}^n$ and $f_1, \ldots, f_p \in \mathbb{R}[x]$ be polynomials such that the set $\{x \in U \mid f_i(x) > 0 \, (i = 1, \ldots, p)\}$ is non-empty. We assume that the product $u(f_1)_+^{s_1} \cdots (f_p)_+^{s_p}$ is well-defined as distribution on $\mathbb{R}^n$ if $s$ belongs to a shift-invariant open set $\Omega$ of $\mathbb{C}^p$. This is the case if $u$ is a locally integrable function. We set $t_i = E_{s_i}$ and $\partial_{t_i} = -s_i E_{s_i}^{-1}$ for $i = 1, \ldots, p$ in the sequel.

**Algorithm 6 (Computing differential equations for $u(f_1)_+^{s_1} \cdots (f_p)_+^{s_p}$)**

**Input:** A set $G_0$ of generators of a holonomic ideal $I$ of $D_n$ annihilating a distribution $u(x)$, and polynomials $f_1, \ldots, f_p \in \mathbb{R}[x]$.
**Output:** A set $G$ of generators of a holonomic ideal $J$ of $D_{n+p}$ annihilating $u(f_1)_+^{s_1} \cdots (f_p)_+^{s_p}$.
1. For $P = P(x, \partial_{x_1}, \ldots, \partial_{x_n}) \in G_0$, set

$$\tau(P) := P\left(x, \partial_{x_1} + \sum_{j=1}^p \frac{\partial f_j}{\partial x_1} \partial_{t_j}, \ldots, \partial_{x_n} + \sum_{j=1}^p \frac{\partial f_j}{\partial x_n} \partial_{t_j}\right).$$

This substitution is well-defined in the ring $D_{n+p}$ since the operators which are substituted for $\partial_{x_1}, \ldots, \partial_{x_n}$ commute with each other.
2. Set

$$G := \{\tau(P) \mid P \in G_0\} \cup \{t_j - f_j(x) \mid j = 1, \ldots, p\}.$$

Now let us prove the correctness of this algorithm:

**Theorem 3** *Let $J$ be the left ideal of $D_{n+p}$ generated by the set*

$$\{\tau(P) \mid P \in I\} \cup \{t_j - f_j(x) \mid j = 1, \ldots, p\}.$$

*Then $J$ is a holonomic ideal of $D_{n+p}$ and annihilates $u(f_1)_+^{s_1} \cdots (f_p)_+^{s_p}$.*

Proof: First note that $J$ is generated by the set

$$\{\tau(P) \mid P \in G_0\} \cup \{t_j - f_j(x) \mid j = 1, \ldots, p\}$$

since $\tau$ is a ring homomorphism. Assume that $u(x)$ is $C^\infty$. Then in view of the equality

$$(\partial_{x_i} u)(f_1)_+^{s_1} \cdots (f_p)_+^{s_p} = \left(\partial_{x_i} + \sum_{j=1}^p \frac{\partial f_j}{\partial x_i} \partial_{t_j}\right)(u(f_1)_+^{s_1} \cdots (f_p)_+^{s_p}),$$

we have
$$\tau(P)(u(f_1)_+^{s_1} \cdots (f_p)_+^{s_p}) = (Pu)(f_1)_+^{s_1} \cdots (f_p)_+^{s_p} = 0 \qquad (6)$$

if $P \in I$ and $s - \nu(1, \ldots, 1) \in \Omega$ for sufficiently large $\nu \in \mathbb{N}$. If $u$ is locally integrable but not $C^\infty$, then (6) makes sense on $\{x \in U \mid f_1(x) \cdots f_p(x) \neq 0\}$ since $(f_1)_+^{s_1} \cdots (f_p)_+^{s_p}$ is $C^\infty$ in $x$ there. We can verify that (6) holds on $U$ by using the same argument as the proof of Lemma 2.9 of [10]. It is easy to see that $t_j - f_j(x)$ annihilates $u(f_1)_+^{s_1} \cdots (f_p)_+^{s_p}$. Hence $J$ annihilates $u(f_1)_+^{s_1} \cdots (f_p)_+^{s_p}$.

Let us show that $D_{n+p}/J$ is holonomic. Since $D_n/I$ is holonomic, its characteristic variety $\mathrm{Char}(D_n/I)$ is an $n$-dimensional algebraic set of $\mathbb{C}^{2n}$. By the definition, we have

$\mathrm{Char}(D_{n+p}/J)$
$$\subset \left\{(x,t,\xi,\tau) \in \mathbb{C}^{2(n+p)} \mid \sigma(P)\left(x, \xi_1 + \sum_{j=1}^{p} \frac{\partial f_j}{\partial x_1}\tau_j, \ldots, \xi_n + \sum_{j=1}^{p} \frac{\partial f_j}{\partial x_n}\tau_j\right) = 0 \quad (\forall P \in I),\right.$$
$$\left. t_j = f_j(x) \ (j=1,\ldots,p)\right\}$$
$$= \left\{(x,t,\xi,\tau) \in \mathbb{C}^{2(n+p)} \mid \left(x, \xi_1 + \sum_{j=1}^{p} \frac{\partial f_j}{\partial x_1}\tau_j, \ldots, \xi_n + \sum_{j=1}^{p} \frac{\partial f_j}{\partial x_n}\tau_j\right) \in \mathrm{Char}(D_n/I),\right.$$
$$\left. t_j = f_j(x) \ (j=1,\ldots,p)\right\}.$$

Since the set on the last line is in one-to-one correspondence with the set $\mathrm{Char}(D_n/I) \times \mathbb{C}^p$, the dimension of $\mathrm{Char}(D_{n+p}/J)$ is $n+p$, which implies that $D_{n+p}/J$ is a holonomic module. $\square$

**Example 10** As one of the simplest examples, let us consider the gamma function
$$v(s) = \Gamma(s+1) = \int_0^\infty e^{-x} x^s \, dx = \int_{-\infty}^\infty e^{-x} x_+^s \, dx.$$
Introducing $t = E_s$ and $\partial_t = -sE_s^{-1}$, we compute a holonomic system
$$(\partial_x + \partial_t + 1)u = (t-x)u = 0$$
for $u(x,s) := e^{-x} x_+^s$. Then integrating with respect to $x$ yields a holonomic system
$$(\partial_t + 1)v(s) = 0$$
for $v(s)$, which can be rewritten as
$$(E_s - (s+1))v(s) = 0.$$

**Example 11** Set
$$v(s) = \int_{x^3 - y^2 \geq 0} e^{-x^2 - y^2}(x^3 - y^2)^s \, dxdy = \int_{\mathbb{R}^2} e^{-x^2 - y^2}(x^3 - y^2)_+^s \, dxdy,$$
which is well-defined for $s$ belonging to the set
$$\Omega := \{s \in \mathbb{C} \mid x \neq -\nu, \ -\frac{5}{6} - \nu, \ -\frac{7}{6} - \nu \quad (\nu = 1, 2, 3, \ldots)\}.$$
The holonomic ideal generated by
$$-x^3 + y^2 + t, \quad \partial_x + 3\partial_t x^2 + 2x, \quad \partial_y + (-2\partial_t + 2)y$$
annihilates the integrand $u(x,s) := e^{-x^2 - y^2}(x^3 - y^2)_+^s$. The integral ideal of this ideal is generated by

$108t^2\partial_t^5 + (-216t^2 + 648t)\partial_t^4 + (108t^2 - 972t + 627)\partial_t^3 + (356t - 606)\partial_t^2 + (-64t + 108)\partial_t + 32t - 48.$

Hence $v(s)$ satisfies a holonomic difference equation

$$(32E_s^4 + (64s + 16)E_s^3 + (-108s^3 + 32s^2 + 32s)E_s^2$$
$$+ (-216s^4 + 540s^3 - 390s^2 + 66s)E_s - 3s(s-1)(s-2)(6s-5)(6s-13))v(s) = 0$$

for $s \in \Omega$.

Differential equations with parameters can be extracted from the output of the integration algorithm as follows.

**Algorithm 7 (intersection with a subring)**

**Input:** A set $G_0$ of generators of a left ideal $J$ of $D_{n+p}$, the ring of differential operators on the variables $(x_1, \ldots, x_n, t_1, \ldots, t_p)$.
**Output:** A set $G$ of generators of the left ideal $J \cap D_n[s]$ of $D_n[s]$ under the identification $s_j = -\partial_{t_j} t_j$ for $j = 1, \ldots, p$.
1. Introducing new variables $u_j, v_j$ for $j = 1, \ldots, p$, let $h(P) \in D_{n+p}[u]$ be the multi-homogenization of $P \in D_{n+p}$; i.e., $h(P)$ is homogeneous with respect to the weight $-1$ for $t_j$ and $u_j$, and $1$ for $\partial_{t_j}$, for each $j$.
2. Let $N$ be the left ideal of $D_{n+p}[u,v]$ generated by the set

$$\{h(P) \mid P \in G_0\} \cup \{1 - u_j v_j \mid j = 1, \ldots, p\}.$$

3. Compute a set $G_1$ of generators of the ideal $N \cap D_{n+p}$ by eliminating $u, v$ via an appropriate Gröbner base.
4. Since each element $P$ of $G_1$ is multi-homogeneous without $u, v$, there exist a monomial $S$ in $t, \partial_t$ and an operator $Q(s) \in D_n[s]$ such that

$$SP = Q(-\partial_{t_1} t_1, \ldots, -\partial_{t_p} t_p).$$

Let $G$ be the set of such $Q$ for each $P \in G_1$.

The correctness of this algorithm was proved as Proposition 4.3 in [19].

**Example 12** Consider the definite integral

$$v(z, s) = \int_{D(t)} (x+t)^s \, dxdy = \int_{\mathbb{R}^2} Y(1 - x^2 - y^2)(x+z)_+^s \, dxdy$$

with $D(z) = \{(x,y) \in \mathbb{R}^2 \mid x^2 + y^2 \leq 1, \ x + z \geq 0\}$, which is well-defined at least for Re $s > 0$ and for any $z \in \mathbb{R}$. We first compute a differential holonomic system for the integrand in the variables $(x, y, z, t)$ with $t = E_s$ and $\partial_t = -sE_s^{-1}$. Then the integration algorithm outputs the ideal generated by

$$\partial_z + \partial_t, \quad (z^2 - 2tz + t^2 - 1)\partial_t + s - t.$$

By computing the intersection with the ring $D_z[s]$ with the identification $s = -\partial_t t$, we get a linear differential equation

$$((1 - z^2)\partial_z^2 + (2s+1)z\partial_z - s(s+2))v(z, s) = 0.$$

# 6 A shortcut for tensor product computation

Let us consider again a definite integral of the form

$$v(x_1, \ldots, x_{n-d}) := \int_{\mathbb{R}^d} u(x) Y(f_1) \cdots Y(f_m) \, dx_{n-d+1} \cdots dx_n$$

for a holonomic function or distribution $u(x)$. In Section 4, we gave an algorithm to compute a holonomic ideal annihilating the integrand by a tensor product computation. Since the tensor product computation can often be a bottleneck in the actual computation, let us describe an alternative method using the technique introduced in the preceding section.

**Algorithm 8 (Computing differential equations for $u(f_1)_+^{\lambda_1} \cdots (f_p)_+^{\lambda_p}$)**

**Input:** A set $G_0$ of generators of a holonomic ideal $I$ of $D_n$ annihilating a distribution $u(x)$, polynomials $f_1, \ldots, f_p \in \mathbb{R}[x]$, and complex numbers $\lambda_1, \ldots, \lambda_p$.
**Output:** A set $G$ of generators of a holonomic ideal of $D_n$ annihilating $u(f_1)_+^{\lambda_1} \cdots (f_p)_+^{\lambda_p}$.
1. Let $G_1$ be the output of Algorithm 6 with input $G_0$ and $\{f_1, \ldots, f_p\}$. Let $J$ be the ideal of $D_{n+p}$ generated by $G_1$.
2. Compute a set $G_2$ of generators of $J \cap D_n[s]$ with the identification $s_j = -\partial_{t_j} t_j$ for $j = 1, \ldots, p$ by using Algorithm 7
3. Set $G := \{P(\lambda) \mid P(s) \in G_2\}$.

**Theorem 4** *The output of the algorithm above generates a holonomic ideal and annihilates $u(f_1)_+^{\lambda_1} \cdots (f_p)_+^{\lambda_p}$ if $u(f_1)_+^{s_1} \cdots (f_p)_+^{s_p}$ is well-defined and holomorphic with respect to $s$ on a neighborhood of $s = \lambda$.*

Proof: It follows from Theorem 3 and the uniqueness of analytic continuation that $G$ annihilates $u(f_1)_+^{\lambda_1} \cdots (f_p)_+^{\lambda_p}$.

Set $M := D_n/I$. Let us show that $J$ coincides with the annihilator of $[1] \otimes f^s$ in $M \otimes_{\mathbb{C}[x]} D_{n+p} f^s$, where $[1]$ denotes the residue class of $1 \in D_n$ in $M$ and $f^s = (f_1)^{s_1} \cdots (f_p)^{s_p}$. Note that $M \otimes_{\mathbb{C}[x]} D_{n+p} f^s$ has a natural structure of left $D_{n+p}$-module.

First, let $P$ be in $J$. Then there exist $Q_1, \ldots, Q_l, \ldots, Q_{l+p} \in D_{n+p}$ and $P_1, \ldots, P_l \in I$ such that

$$P = \sum_{i=1}^{l} Q_i \tau(P_i) + \sum_{j=1}^{p} Q_{l+j} \cdot (t_j - f_j(x)).$$

It follows that

$$P([1] \otimes f^s) = \sum_{i=1}^{l} Q_i([P_i] \otimes f^s) + \sum_{j=1}^{p} Q_{l+j}([1] \otimes (t_j - f_j(x)) f^s) = 0.$$

Conversely, suppose that $P \in D_{n+p}$ annihilates $[1] \otimes f^s$. We can rewrite $P$ in the form

$$P = \sum_{\alpha \in \mathbb{N}^n, \nu \in \mathbb{N}^p} p_{\alpha,\nu}(x)\Big(\partial_{x_1} + \sum_{j=1}^{p} \frac{\partial f_j}{\partial x_1}\partial_{t_j}\Big)^{\alpha_1} \cdots \Big(\partial_{x_n} + \sum_{j=1}^{p} \frac{\partial f_j}{\partial x_n}\partial_{t_j}\Big)^{\alpha_n} \partial_t^\nu$$

$$+ \sum_{j=1}^{p} Q_j \cdot (t_j - f_j(x))$$

with $p_{\alpha,\nu}(x) \in \mathbb{C}[x]$ and $Q_j \in D_{n+p}$. Setting $P_\nu := \sum_{\alpha \in \mathbb{N}^n} p_{\alpha,\nu}(x)\partial_x^\alpha$, we get

$$0 = P([1] \otimes f^s) = \sum_{\nu \in \mathbb{N}^p} [P_\nu] \otimes (\partial_t^\nu f^s) \in M \otimes_{\mathbb{C}[x]} D_{n+p}f^s.$$

It follows that each $P_\nu$ belongs to $I$ since $\{\partial_t^\nu f^s\}$ constitute a free basis of $D_{n+p}f^s$ over $\mathbb{C}[x]$ (See e.g., Lemma 6.11 of [15]). Hence we have

$$P = \sum_{\nu \in \mathbb{N}^p} \partial_t^\nu \tau(P_\nu) + \sum_{j=1}^{p} Q_j \cdot (t_j - f_j(x)) \in J.$$

There is a natural homomorphism

$$\rho : M \otimes_{\mathbb{C}[x]} D_n[s]f^s \to M \otimes_{\mathbb{C}[x]} D_{n+p}f^s \simeq D_{n+p}/J$$

of left $D_n[s]$-module which sends $[1] \otimes f^s$ to the modulo class $[1]$ in $D_{n+p}/J$. Moreover we have

$$\rho(M \otimes_{\mathbb{C}[x]} D_n[s]f^s) = D_n[s]/(J \cap D_n[s]) \subset D_{n+p}/J.$$

Thus $\rho$ induces a homomorphism

$$\rho' : M \otimes_{\mathbb{C}[x]} D_n f^\lambda \to D_n[s]/\Big((J \cap D_n[s]) + \sum_{j=1}^{p}(s_j - \lambda_j)D_n[s]\Big)$$

of left $D_n$-module, which sends $[1] \otimes f^\lambda$ to the residue class of $1$, where $f^\lambda$ denotes the residue class of $f^s$ in $D_n[s]f^s/(\sum_{j=1}^{p}(s_j - \lambda_j)D_n[s]f^s)$. Let $N$ be the left ideal of $D_n$ generated by the output of Algorithm 8. Then by the definition we have an isomorphism

$$D_n/N \simeq D_n[s]/\Big((J \cap D_n[s]) + \sum_{j=1}^{p}(s_j - \lambda_j)D_n[s]\Big)$$

as $D_n$-module. In conclusion, $\rho'$ can be identified with a surjective homomorphism

$$\rho' : M \otimes_{\mathbb{C}[x]} D_n f^\lambda \to D_n/N.$$

This implies that $D_n/N$ is holonomic in view of Proposition 1. $\square$

**Remark 1** The homomorphisms $\rho$ or $\rho'$ are not injective in general; for example, set $f = x_1 x_2$ and $M := D_n/I$ with $I$ being the left ideal of $D_2$ generated by $x_1$ and $\partial_2$. Then

$$\partial_2([1] \otimes f^s) = [1] \otimes \partial_2 f^s = -[1] \otimes x_1 \partial_t f^s = -[x_1] \otimes \partial_t f^s = 0$$

holds in $M \otimes_{\mathbb{C}[x]} D_2 \langle t, \partial_t \rangle f^s$ but not in $M \otimes_{\mathbb{C}[x]} D_2[s] f^s$. The annihilator of $[1] \otimes f^0$ in $M \otimes_{\mathbb{C}[x]} D_2 f^0$ is generated by $x_1$ and $x_2 \partial_2$, while $N$ for $\lambda = 0$ is generated by $x_1$ and $\partial_2$.

**Example 13** The Bessel function $J_\nu(x)$ satisfies a holonomic difference-differential system

$$(x^2 \partial_x^2 + x \partial_x + x^2 - \nu^2) J_\nu(x) = (x(E_\nu^2 + 1) - 2(\nu + 1) E_\nu) J_\nu(x) = 0 \quad (7)$$

for $x \in \mathbb{R}$ and $\nu \in \mathbb{C} \setminus \{-1, -2, \ldots\}$ and holomorphic in $\nu$. Let us consider the integral

$$v(z, \nu, s) := \int_{D(z)} y^s J_\nu(x) \, dx dy$$

with $D(z) := \{(x, y) \in \mathbb{R}^2 \mid x, y \geq 0, x^2 + y^2 \leq z\}$. It is easy to verify that $Y(x) J_\nu(x)$ also satisfies (7). First we compute a holonomic system for

$$u(x, y, z, \nu, s) := Y(z - x^2 - y^2) y_+^s Y(x) J_\nu(x)$$

in the ring of differential operators on the variables $x, y, z, t_1, t_2$ with $t_1 = E_\nu$ and $t_2 = E_s$ by using Algorithm 8. Tensor product computation fails for this example because of complexity. Then by the integration algorithm, we get a holonomic system for $v(z, \nu, s)$ in the ring of differential operators on the variables $z, \nu, s$. Finally computing the intersection with the subring $D_z[\nu, s]$ we get

$$(8z^2 \partial_z^3 + (-8s + 8) z \partial_z^2 + (2z - 2\nu^2 + 2s^2) \partial_z - s - 1) v(z, \nu, s) = 0,$$

which has a regular singularity at $t = 0$ with exponents $1, (\pm \nu + s)/2 + 1$. This differential equation is valid, at least, for $z \in \mathbb{R}$, $\operatorname{Re} s > -1$, and $\nu \neq -1, -2, \ldots$.

Finally we present timing data (in seconds) of the computation of difference equations in $\nu$ for integrals involving $J_\nu(x)$ by using Risa/Asir running on a computer with 3.06 GHz Core 2 Duo Processor and 8 Gbyte memory. Step 1 refers to the computation of a holonomic system for the integrand by using Algorithm 8; Step 2 consists of the integration algorithm (Algorithm 3). Tensor product algorithm does not stop in a reasonable time period for most of the examples in the table.

| integral | Step 1 | Step 2 | total |
|---|---|---|---|
| $\int Y(y) Y(1 - x - y) Y(x) J_\nu(x) \, dx dy$ | 0.08 | 0.03 | 0.11 |
| $\int Y(y) Y(1 - x^2 - y^2) Y(x) J_\nu(x) \, dx dy$ | 0.2 | 2.2 | 2.4 |
| $\int Y(y) Y(1 - x^4 - y^4) Y(x) J_\nu(x) \, dx dy$ | 3.4 | 346 | 349 |
| $\int Y(y) Y(z) Y(1 - x - y - z) Y(x) J_\nu(x) \, dx dy dz$ | 0.27 | 0.13 | 0.4 |
| $\int Y(y) Y(z) Y(1 - x^2 - y^2 - z^2) Y(x) J_\nu(x) \, dx dy dz$ | 0.13 | 2.4 | 2.5 |

**Acknowledgement** This work was supported by JSPS Grant-in-Aid for Scientific Research (C) 21540200.

# References

[1] Almkvist, G., Zeilberger, D., The method of differentiating under the integral sign. J. Symbolic Computation **10** (1990), 571–591.

[2] Bernstein, J., Modules over a ring of differential operators. An investigation of the fundamental solutions of equations with constant coefficients. (Russian) Funkcional. Anal. i Prilozen. **5** (1972), 1–16.

[3] Björk, J.-E., Rings of Differential Operators. North-Holland, 1979.

[4] Briançon, J., Maisonobe, Ph., Remarques sur l'idéal de Bernstein associé à des polynômes. Preprint Université de Nice Sophia-Antipolis, no. 650, 2002.

[5] Chyzak, F., Fonctions holonomes en calcul formel. Thèse de doctorat, École polytechnique–INRIA, TU0531, 227 pages.

[6] Chyzak, F and Salvy, B., Non-commutative elimination in Ore algebras proves multivariate identities. Journal of Symbolic Computation, **26** (1998), 187–227.

[7] Gel'fand, I. M.; Shilov, G. E. Generalized functions. Vol. 1. Properties and operations. Translated from the Russian by Eugene Saletan. Academic Press, New York, London, 1964.

[8] Gyoja, A., Bernstein-Sato's polynomial for several analytic functions, J. Mathematics Kyoto University. **33** (1993), 399–411.

[9] Kashiwara, M., *B*-functions and holonomic systems — rationality of roots of *b*-functions. Invent. Math. **38** (1976), 33–53.

[10] Kashiwara, M., Kawai, T., On the characteristic variety of a holonomic system with regular singularities. Adv. Math. **34** (1979), 163–184.

[11] Nakayama, H., Nishiyama, K., An algorithm of computing inhomogeneous differential equations for definite integrals. ICMS'10 Proceedings of the Third international conference on Mathematical software, Springer-Verlag Berlin, Heidelberg, 2010.

[12] Noro, M., Takayama, N., Nakayama, H., Nishiyama, K., Ohara, K, Risa/Asir: a computer algebra system. http://www.math.kobe-u.ac.jp/Asir/asir.html, 2011.

[13] Oaku, T., Computation of the characteristic variety and the singular locus of a system of differential equations with polynomial coefficients. Japan J. Indust. Appl. Math. **11** (1994), 485–497.

[14] Oaku, T., An algorithm of computing *b*-functions. Duke Math. J. **87** (1997), 115–132.

[15] Oaku, T., Algorithms for *b*-functions, restrictions, and algebraic local cohomology groups of *D*-modules. Advances in Appl. Math. **19** (1997), 61–105.

[16] Oaku, T., Regular *b*-functions of *D*-modules. J. Pure Appl. Algebra **213** (2009), 1545–1557.

[17] Oaku, T., Shiraki, Y., Takayama, N., Algebraic algorithms for $D$-modules and numerical analysis. Computer mathematics (Proceedings of ASCM 2003), 23–39, Lecture Notes Ser. Comput., 10, World Sci. Publ., River Edge, NJ, 2003.

[18] Oaku, T., Takayama, N., An algorithm for de Rham cohomology groups of the complement of an affine variety. J. Pure Appl. Algebra **139** (1999), 201–233.

[19] Oaku, T., Takayama, N., Algorithms for $D$-modules — restriction, tensor product, localization, and local cohomology groups. J. Pure Appl. Algebra **156** (2001), 267–308.

[20] Sato, M., Kawai, T., Kashiwara M., Microfunctions and pseudo-differential equations. Springer Lecture Notes in Math. No. 287 (1973), pp. 264–529.

[21] Saito, M., Sturmfels, B., Takayama, N., Gröbner Deformations of Hypergeometric Differential Equations. Springer Verlag, 2000.

[22] Schwartz, L., Théorie des distributions. Hermann, 1966, Paris.

[23] Takayama, N., An algorithm of computing the integral of a module — an infinite dimensional analog of Gröbner basis. In: Symbolic and Algebraic Computation, Proceedings of ISSAC '90, Kyoto, S. Watanabe and M. Nagata, eds., pp. 206–211, ACM and Addison-Wesley, 1990.

[24] Takayama, N., An approach to the zero recognition problem by Buchberger algorithm. J. Symbolic Computation **14** (1992), 265–282.